\documentclass[a4paper,11pt,notitlepage]{article}
\usepackage[notcite,notref]{}
\usepackage{latexsym}
\usepackage{amssymb}
\usepackage{color}
\usepackage{graphicx}
\usepackage{amsmath}
\usepackage{amsthm}
\usepackage{mathrsfs}
\usepackage{natbib}
\usepackage{amsfonts}
\usepackage{accents}
\usepackage{enumitem}
\usepackage{stmaryrd}     % for \llbrackets
\usepackage{subeqnarray}
\usepackage{subfigure}
\usepackage{float} 
\usepackage{bbm}
\usepackage{hyperref}
\usepackage[ruled,vlined]{algorithm2e}
\usepackage[table]{xcolor}
\usepackage{endnotes}
\usepackage{array}
\usepackage{abstract}
\usepackage{multirow}
\usepackage{chngpage}
\usepackage{url}
\usepackage[hang,font=scriptsize]{caption}
\usepackage{float}
\usepackage{cases}
\newtheorem{theorem}{Theorem}[section]

\newtheorem{lemma}[theorem]{Lemma}
\newtheorem{remark}[theorem]{Remark}

\newtheorem{prop}[theorem]{Proposition}

\DeclareMathOperator*{\argmax}{argmax}

\def \Frac{\displaystyle\frac}

\def \Sup{\displaystyle\sup}

\def \Maxs#1#2 {\displaystyle\max_{\substack{#1 \\ #2}}}
\def \Sups#1#2 {\displaystyle\sup_{\substack{#1 \\ #2}}}

\def \dot#1#2#3{#1 = #2,\; \dots,\; #3}

\newcommand{\al}{\alpha}

\definecolor{myDarkGrey}{RGB}{132, 132, 132}
\definecolor{myMidGrey}{RGB}{190, 190, 190}
\definecolor{myLightGrey}{RGB}{216, 216, 216}
\definecolor{myWhite}{RGB}{255,255,255}
\definecolor{myAzure}{rgb}{0.60784314,0.89803922,1}

\def\hB#1{\hat{B}_{#1}}

\newcommand{\covB}{\Sigma_0}
\newcommand{\covE}{\Gamma}
\newcommand{\ivCovE}{\covE^{-1}}

\def\X#1{X^{\al}_{#1}}

\def\Z#1{Z^{\al}_{#1}}

\def\A{\mathcal{A}}
\def\Sc{\mathcal{S}}

\def\Aq#1{\A^q_{#1}}

\def\N{\mathbb N}
\def\n{\nu}
\def\m#1{\mu_{#1}}
\def\P{\mathbb P}

\def\E{\mathbb E}
\def\Fb{\mathbb F}
\def\Gb{\mathbb G}

\def\e#1{\epsilon_{#1}}
\def\te#1{\tilde{\epsilon}_{#1}}
\def\P{\mathbb P}
\def\F#1{\mathcal F_{#1}}
\def\R{\mathbb R}
\def\Rd{\mathbb R^d}
\def\ind{\perp\!\!\!\perp}

\usepackage[a4paper,top=3cm,bottom=3cm,left=2cm,right=3cm,showframe=False]{geometry}

\def \ep{\hbox{ }\hfill$\Box$}
\def \epR{\hbox{ }\hfill$\lozenge$}

\def\beqs{\begin{eqnarray*}}
\def\enqs{\end{eqnarray*}}
\def\beq{\begin{eqnarray}}
\def\enq{\end{eqnarray}}

\def\beqal{\beq \begin{aligned}}
\def\enqal{\end{aligned} \enq }
\def\beqsal{\beqs \begin{aligned}}
\def\enqsal{\end{aligned} \enqs }

\def\beqalac{\beq \left \{ \begin{aligned}}
\def\enqalac{\end{aligned} \right. \enq }

\def\beqsalac{\beqs \left \{ \begin{aligned}}
\def\enqsalac{\end{aligned} \right. \enqs}

\SetKwInput{KwInput}{Input}                % Set the Input
\SetKwInput{KwOutput}{Output}              % set the Output
\SetKwComment{Comment}{$\triangleright$\ }{}

\SetCommentSty{mycommfont}

\title{Discrete-time portfolio optimization under maximum drawdown constraint with partial information and deep learning resolution} %\footnote{This work is issued from a CIFRE collaboration between OSSIAM and LPSM.}}
\author{Carmine DE FRANCO\footnote{OSSIAM, E-mail:
    carmine.de-franco@ossiam.com}\and Johann NICOLLE\footnote{OSSIAM and LPSM, Universit\'e de Paris, E-mail:
    johann.nicolle@ossiam.com}\and Huy\^en PHAM \footnote{LPSM, Universit\'e de Paris, E-mail: pham@lpsm.paris}}
\begin{document}

\frenchspacing

\maketitle

\begin{abstract}
We study a discrete-time portfolio selection  problem with partial information and  maxi\-mum drawdown constraint. Drift uncertainty in the multidimensional framework is modeled by a prior probability distribution. In this Bayesian framework, we derive the dynamic programming equation using an appropriate change of measure, and obtain semi-explicit results in the Gaussian case. The latter case, with a CRRA utility function is completely solved numerically  using recent deep learning techniques for stochastic optimal control problems. 
We emphasize the informative value of the learning strategy versus the non-learning one by providing empirical performance and sensitivity analysis with respect to the uncertainty of the 
drift.  Furthermore, we show numerical evidence of the close relationship between the non-learning strategy and a no short-sale constrained Merton problem, by 
illustrating the convergence of the former towards the latter as the maximum drawdown constraint vanishes.
\end{abstract}

%%%%%%%%%%%%%%%%%%%%%%%%%%%%%%%%%%%%%%%%%%%%%%%%%%%%%%%%%%%%%%%%%%%%%%%%%%%%%%%%%%%%%%%%%%%%%%%%%%%%%%%%%%%%%%%%%%%%%%%%%%%%%%%%%%%%%%%%%%%%%%%%%%%%%%%%%%%%%%%%%%%%%%%%%%%%%%%%%%%%%%%%%%%%%%%%%%%%%%%%%%%%%%%%%%%%%%%%%%%%%%%%
%Introduction
%%%%%%%%%%%%%%%%%%%%%%%%%%%%%%%%%%%%%%%%%%%%%%%%%%%%%%%%%%%%%%%%%%%%%%%%%%%%%%%%%%%%%%%%%%%%%%%%%%%%%%%%%%%%%%%%%%%%%%%%%%%%%%%%%%%%%%%%%%%%%%%%%%%%%%%%%%%%%%%%%%%%%%%%%%%%%%%%%%%%%%%%%%%%%%%%%%%%%%%%%%%%%%%%%%%%%%%%%%%%%%%%
\section{Introduction}
 
This paper is devoted to the study of a constrained allocation problem in discrete time with partial information. We consider an investor who is willing to maximize the expected utility of her terminal wealth over a given investment horizon. The risk-averse investor is looking for the optimal portfolio in financial assets under a maximum drawdown constraint. The maximum drawdown is a common metric in finance and represents the largest drop in the portfolio value. Our framework incorporates this constraint by 
setting a threshold representing the proportion of the current maximum of the wealth process that the investor is willing to keep.

The expected rate of  assets' return (drift) is unknown, but information can be learnt  by progressive observation of the financial asset prices. 
The uncertainty about the rate of return  is modeled by a probability distribution, i.e., a prior belief on the drift. 
%dealt with, by assuming the investor has a view on the probability distribution of the drift, i.e. a prior. 
To take into account the information conveyed by the prices, this prior will be updated using a Bayesian learning approach. \\

An extensive literature exists on parameters uncertainty and especially on filtering and learning techniques in a partial information framework. To cite just a few, see \cite{Lakner1998}, \cite{Rogers2001}, \cite{Cvitanic2006}, \cite{Karatzas2001}, \cite{Gueant2017}, and \cite*{dfjnhp19BL}. 
Some articles deal with risk constraints in a portfolio allocation framework, see for instance the paper by \cite{Redeker2018} which tackles dynamic risk constraints and compares the continuous and discrete time trading. Other papers especially focus on drawdown constraints, see in particular the seminal paper by \cite{Grossman1993} or \cite{Cvitanic1994}. More recently, \cite{Elie2008} study  infinite-horizon optimal consumption-investment problem in continuous-time, and \cite{Boyd2019} use forecasts of the mean and covariance of financial returns from a multivariate hidden Markov model with time-varying parameters to build the optimal controls. 

As it is not possible to solve analytically our constrained optimal allocation problem, we have applied a machine learning algorithm  developed in \cite{Bachouch2018a} and 
\cite{Bachouch2018b}. This algorithm, called \emph{Hybrid-Now}, is particularly suited for solving stochastic control problems in high dimension using deep neural networks.

Our main contributions to the literature is twofold: a detailed theoretical study of a discrete-time portfolio selection problem including both drift uncertainty and maximum drawdown constraint, and a numerical resolution using a deep learning approach for an application to a model of three risky assets, leading to a five-dimensional problem. We derive  the dynamic programming equation (DPE), which is in general of infinite-dimensional nature, following the change of measure suggested in \cite{Elliott2008}. In the Gaussian case, the DPE is reduced to a finite-dimensional equation by exploiting the Kalman filter.  
In the particular case of constant relative risk aversion (CRRA) utility function, we  reduce furthermore the dimensionality of the problem. 
Then, we solve numerically the problem in the Gaussian case with CRRA utility functions using the deep learning \emph{Hybrid-Now} algorithm. 
Such numerical results allow us to provide a detailed analysis of the performance and allocations of both the learning and non-learning strategies benchmarked with a comparable equally-weighted strategy. Finally, we assess the performance of the learning compared to the non-learning strategy with respect to the sensitivity of the uncertainty of the drift. Additionally, we provide  empirical evidence of convergence of the non-learning strategy to the solution of the classical Merton problem when the parameter controlling the maximum drawdown vanishes.
 
The paper is organized as follows: Section \ref{sec: Pb_Setup} sets up the financial market model and the asso\-ciated optimization problem. Section \ref{sec: Gen_case} describes, in the general case, the change of measure and the Bayesian filtering, the derivation of the dynamic programming equation and details some properties of the value function. Section \ref{sec: Gauss_case} focuses on the Gaussian case. Finally, Section \ref{sec: Deep learning numerical resolution} presents the neural network techniques used, and shows the numerical results.
 
%%%%%%%%%%%%%%%%%%%%%%%%%%%%%%%%%%%%%%%%%%%%%%%%%%%%%%%%%%%%%%%%%%%%%%%%%%%%%%%%%%%%%%%%%%%%%%%%%%%%
%Problem Setup
%%%%%%%%%%%%%%%%%%%%%%%%%%%%%%%%%%%%%%%%%%%%%%%%%%%%%%%%%%%%%%%%%%%%%%%%%%%%%%%%%%%%%%%%%%%%%%%%%%%%
\section{Problem setup} \label{sec: Pb_Setup}

On a probability space $(\Omega, \F{}, \P)$ equipped with a discrete filtration $(\F{k})_{\dot{k}{0}{N}}$ satisfying the usual conditions, we consider a financial market model with  
one riskless asset assumed normalized to one, and $d$ risky assets.  The price process $(S^i_k)_{k=0,...,N}$ of asset  $i \in [\![1,d]\!]$ is governed by the dynamics
\beq \label{eq: price_dyn}
			S^i_{k+1} & = \; S^i_k e^{R^i_{k+1}}, \quad k = 0,\ldots,N-1, 
\enq
where $R_{k+1}$ $=$ $(R^1_{k+1},\ldots,R^N_{k+1})$ is the vector of the assets log-return between time $k$ and $k+1$, and modeled as:
\beq \label{eq: ret_process}
R_{k+1} & = \; B^{} + \e{k+1}.
\enq
The drift vector $B$ is a $d$-dimensional random variable with probability distribution (prior) $\m{0}$ of known mean $b_0$ $=$ $\E[B]$ and finite second order moment. 
Note that the case of known drift  $B$ means that  $\mu_0$ is a Dirac distribution. The noise $\e{}$ $=$ $(\e{k})_k$ is a sequence of centered  i.i.d. random vector variables with 
covariance matrix $\covE$ $=$ $\E[\e{k} \e{k}^\prime]$, and assumed to be independent of $B$.  
We also assume the fundamental assumption that the probability distribution $\n$ of $\e{k}$ admits a strictly positive density function $g$ on $\R^d$ with respect to the Lebesgue measure.

The price process $S$ is observable, and notice by relation  \eqref{eq: price_dyn} that $R$ can be deduced from $S$, and vice-versa. 
We will then denote by  $\Fb^o$ $=$ $\left\{ \F{k}^o \right\}_{\dot{k}{0}{N}}$ the observation filtration generated by the process $S$ (hence equivalently by $R$) 
 augmented by the null sets of $\F{}$, with the convention that for $k$ $=$ $0$, $\F{0}^o$ is the trivial algebra. 

An investment strategy is an $\Fb^o$-progressively measurable process $\al = (\al_k)_{\dot{k}{0}{N-1}}$, valued in $\R^d$, and 
representing the proportion of the current wealth invested in each of the $d$ risky assets at each time $\dot{k}{0}{N-1}$. 
Given an investment strategy $\al$ and an initial wealth $x_0 >0$, the (self-financed) wealth process $X^{\al}$ evolves according to 
\beqalac \label{eq: return_wealth}
			\X{k+1} &=  \X{k} \left( 1 + \al_k^\prime \left(e^{R_{k+1}}- \mathbbm{1}_d \right) \right), \quad  \quad \dot{k}{0}{N-1},\\
			\X{0} & =x_0.
\enqalac
where $e^{R_{k+1}}$ is  the $d$-dimensional random variable with components $\left[ e^{R_{k+1}} \right]_i = e^{R^i_{k+1}}$ for $i \in [\![1,d]\!]$, 
and $\mathbbm{1}_d$ is the vector in $\R^d$ with all components equal to $1$. 

Let us  introduce the process $\Z{k}$, as the maximum up to time $k$ of the wealth process $\X{}$, i.e., 
\beqs
	\Z{k} & := &  \max_{0 \leq \ell  \leq k}\X{\ell}, \quad \quad \dot{k}{0}{N}. 
\enqs
The maximum drawdown constraints the wealth $\X{k}$ to remain above a fraction $q \in (0,1)$ of the current historical maximum $\Z{k}$.  
We then define the set of {\it admissible} investment strategies $\Aq{0}$ as the set of investment strategies $\alpha$ such that 
\beqs
\X{k}  & \geq & q  \Z{k},  \quad  {\rm a.s}., \quad k  = 0,\ldots,N. 
\enqs
 
In this framework, the portfolio selection problem is formulated as 
\beq \label{eq: pb_init}
V_0 \; := \; \sup_{\alpha \in \Aq{0}}\E \left[ U\left( \X{N}\right)\right],
\enq
where $U$ is a utility function on $(0,\infty)$ satisfying the standard Inada conditions: continuously differentiable, strictly increasing, concave on $(0, \infty)$ with $U'(0) = \infty$ and $U'(\infty) = 0$.

%%%%%%%%%%%%%%%%%%%%%%%%%%%%%%%%%%%%%%%%%%%%%%%%%%%%%%%%%%%%%%%%%%%%%%%%%%%%%%%%%%%%%%%%%%%%%%%%%%%%
%Dynamic programming system
%%%%%%%%%%%%%%%%%%%%%%%%%%%%%%%%%%%%%%%%%%%%%%%%%%%%%%%%%%%%%%%%%%%%%%%%%%%%%%%%%%%%%%%%%%%%%%%%%%%%
\section{Dynamic programming system} \label{sec: Gen_case}
In this section, we show how Problem \eqref{eq: pb_init} can be characterized from dynamic programming in terms of a backward system of equations amenable for algorithms. In a first step, we will update the prior on the drift uncertainty,  and take advantage of the newest available information by adopting a Bayesian filtering approach. This relies on a suitable 
change of probability measure. 

\subsection{Change of measure and Bayesian filtering} \label{subsec: Chg_meas}
We start by introducing a change of measure under which $R_1$,..., $R_N$ are mutually independent, identically distributed random variables and independent from the drift $B$, hence behaving like a noise.  Following the methodology detailed in \citep{Elliott2008} we define the $\sigma$-algebras 
\beqs
	\mathcal{G}^0_k := \sigma(B,R_1, \dots, R_k), \quad k=0,\ldots,N, 
\enqs
and $\Gb$ $=$ $(\mathcal{G}_k)_k$ the corresponding complete filtration. We then define a new probability measure $\overline\P$ on $(\Omega,\bigvee_{k=1}^N \mathcal{G}_k)$ by 
\beqs
\frac{{\rm d}\overline\P}{{\rm d} \P}\bigg|_{\mathcal{G}_{k}} &:=&  \Lambda_k, \quad k=0,\ldots,N, 
\enqs
with
\beqs
\Lambda_k &:=&  
\prod_{\ell=1}^k \frac{g(R_\ell)}{g(\e{\ell})}, \quad k=1,\ldots,N, \;\;  \Lambda_0 \; = \; 1.  
\enqs
The  existence of $\overline\P$  comes from the Kolmogorov's theorem since $\Lambda_k$ is a strictly positive martingale with expectation equal to one. Indeed, for all $k= 1,..., N$,
\begin{itemize}
\item $\Lambda_k > 0$ since the probability density function  $g$ is strictly positive 
\item $\Lambda_k$ is $\mathcal{G}_k$-adapted,
\item As  $\e{k} \ind \mathcal{G}_{k-1}$, we have 
\beqs
\E[ \Lambda_k|\mathcal{G}_{k-1}] &=& 
 \Lambda_{k-1} \E\Big[ \frac{g(B+\e{k})}{g(\e{k})} \big | \mathcal{G}_{k-1}\Big] \\
			& =&  \Lambda_{k-1} \int_{\Rd} \frac{g(B+ e )}{g(e)} g(e)de  \; =  \; \Lambda_{k-1} \int_{\Rd} g(z)dz  \; = \;  \Lambda_{k-1}. 
\enqs 
\end{itemize}

\begin{prop} \label{lem: prop_p_bar}
Under $\overline \P$, $\left(R_k\right)_{k=1,\dots, N}$, is a sequence of i.i.d. random variables, independent from $B$, having the same probability distribution $\nu$ as $\e{k}$.
\end{prop}
\noindent {\bf Proof.}
See Appendix \ref{Pr: lem: prop_p_bar}.
\ep

Conversely, we recover the initial measure $\P$ under which $(\e{k})_{k = 1, \dots, N}$ is a sequence of independent and identically distributed random variables having probability density function $g$ where $\e{k} = R_k-B$. Denoting by  $\overline{\Lambda}_k$ the Radon-Nikodym derivative ${\rm d}\P/ {\rm d}\overline \P$ restricted to the $\sigma $-algebra $\mathcal{G}_{k}$: 
\beqs
	\frac{{\rm d} \P}{{\rm d}\overline\P}\bigg|_{\mathcal{G}_{k}} = \overline{\Lambda}_k,
\enqs
we have
\beqs
	\overline{\Lambda}_k \; = \;  
\prod_{i=1}^k \frac{g(R_i-B)}{g(R_i)}.
\enqs
It is clear that, under $\P$, the return and wealth processes have the form stated in equations \eqref{eq: ret_process} and \eqref{eq: return_wealth}.
Moreover, from Bayes formula, the posterior distribution of the drift, i.e. the conditional law of $B$ given the asset price observation, is 
\beq \label{defmuk} 
\mu_k(db) \; := \; \P\big[ B \in db  | \F{k}^o  \big] &=& \frac{\pi_k(db)}{\pi_k(\R^d)}, 
\quad k=0,\ldots,N,
\enq
where  $\pi_k$ is the so-called unnormalized conditional law 
\beqs \label{eq: unorm_cond_distri}
\pi_k(db):=\overline{\E} \big[ \overline{\Lambda}_k \mathbbm{1}_{\{B \in db\}} | \F{k}^o \big], \quad k=0,\ldots,N. 
\enqs

We then have the key  recurrence linear relation on the unnormalized conditional law. 

\begin{prop} \label{lem: unnorm_cond_law}
We have the recursive linear  relation   
 \beq \label{recurpi}
\pi_\ell &=& \bar g(R_\ell - \cdot) \pi_{\ell-1},   \quad \ell =1,\ldots,N, 
\enq
with initial condition $\pi_0$ $=$ $\mu_0$, 
where 
\beqs
\bar g(R_\ell - b) &=& \frac{g(R_{\ell}-b)}{g(R_{\ell})}, \quad b \in \R^d, 
\enqs
and we recall that $g$ is the probability density function of the identically distributed $\e{k}$ under $\P$.
\end{prop}
\noindent {\bf Proof.}
See Appendix \ref{Pr: lem: unnorm_cond_law} .
\ep

\subsection{The static set of admissible controls} \label{subsec: space_controls}
In this subsection, we derive some useful characteristics of the space of controls which will turn out to be crucial in the derivation of  the dynamic programming system.

Given  time $k$ $\in$ $[\![0,N]\!]$,  a current wealth $x$ $=$ $X_k^\alpha$ $>$ $0$, and current maximum wealth $z$ $=$ $Z_{k}^\alpha$ $\geq$ $x$ 
 that satisfies the drawdown constraint $qz$ $\leq$ $x$ at time $k$ for an admissible investment strategy $\alpha$ $\in$ $\Aq{0}$, we denote by $A_k^q(x,z)$ $\subset$ $\R^d$  the set of static controls $a$ $=$ $\alpha_k$ such that the drawdown constraint is satisfied at next time $k+1$, i.e. $X_{k+1}^{\al}$ $\geq$ $q Z_{k+1}^{\al}$. From the relation \eqref{eq: return_wealth}, and noting that 
$\Z{k+1}$ $=$ $\max[\Z{k},\X{k+1}]$,  this yields
 \beqal \label{Aqxz}
&A^q_{k}(x,z) =  \\
& \hspace{5mm}  \left \{ a \in \Rd : 1 +  a^\prime \big(e^{R_{k+1}}- \mathbbm{1}_d \big)
\; \geq \; q \max\Big[ \frac{z}{x}, 1 +  a^\prime \big(e^{R_{k+1}}- \mathbbm{1}_d \big)  \Big]\; {\rm a.s.} \right \}.
\enqal
Recalling from Proposition \ref{lem: prop_p_bar}, that the random variables $R_{1},...,R_{N}$ are i.i.d. under $\overline \P$, we notice that the set $A_k^q(x,z)$ does not depend on the current time $k$, and we will drop the subscript $k$ in the sequel, and simply denote by $A^q(x,z)$.  

Remembering that the support of $\nu$, the probability distribution of $\e{k}$, is $\Rd$, the following lemma characterizes more precisely the set $A^q(x,z)$.

\begin{lemma} \label{lem: cond_a}
For any $(x,z)$ $\in$ $\Sc^q$ $:=$ $\big\{ (x,z) \in (0,\infty)^2: qz \leq x \leq z \big\}$, we have 
\beqs
A^q{(x,z)} &=& \Big\{ a \in \Rd_+ : |a|_{_1} \leq 1-q\frac{z}{x} \Big\}, 
\enqs
where $|a|_{_1}$ $=$ $\sum_{i=1}^d |a_i|$ for $a$ $=$ $(a_1,\ldots,a_d)$ $\in$ $\R^d_+$. 
\end{lemma}
\noindent {\bf Proof.}
See Appendix \ref{Pr: lem: cond_a}.
\ep

Let us prove some properties on the admissible set $A^q(x,z)$.

\begin{lemma} \label{lem: Aq_gen}
For any $(x,z)$ $\in$ $\Sc^q$,  the set $A^q{(x,z)}$ satisfies the following properties:
\begin{enumerate}
	\item It is decreasing in $q$: $\forall q_1 \leq q_2,$ $A^{q_2}{(x,z)} \subseteq	 A^{q_1}{(x,z)}$, \label{lem: Aq_decreasing_q__gen} 
	\item It is continuous in $q$, \label{lem: Aq_continuous_q__gen} 
	\item It is increasing in $x$: $\forall x_1 \leq x_2,$ $A^q{(x_1,z)} \subseteq	 A^{q}{(x_2,z)}$, \label{lem: Aq_increasing_x_gen}
	\item It is a convex set, \label{lem: Aq__convex_gen}
	\item It is homogeneous: $a$ $\in$ $A^q{}(x,z)$ $\Leftrightarrow$ $a$ $\in$ $A^q{}(\lambda x, \lambda z)$, for any $\lambda$ $>$ $0$.  
\end{enumerate}
\end{lemma}
\noindent {\bf Proof.}
See Appendix \ref{Pr: lem: Aq_gen}.
\ep

\subsection{Derivation of the dynamic programming equation} \label{subsec: Dyn_prog_eq}
The change of probability detailed in Subsection \ref{subsec: Chg_meas} allows us to turn the initial partial information Problem \eqref{eq: pb_init} into a 
full observation problem as 
\beq 
V_0 \; := \; \sup_{\al \in \Aq{0}} \E[U(\X{N})] &=&  \sup_{\al \in \Aq{0}} \overline\E[ \overline{\Lambda}_N U(\X{N})] \nonumber \\
&=& \sup_{\al \in \Aq{0}} \overline\E \Big[ \overline\E \big[ \overline{\Lambda}_N U(\X{N}) \big| \F{N}^o \big] \Big] \nonumber \\
&=& \sup_{\al \in \Aq{0}} \overline\E \Big[ U(\X{N})  \pi_N(\R^d) \Big],  \label{eq: problem_U change}
\enq
from Bayes formula, the law of conditional expectations, and the definition of the unnormalized filter $\pi_N$ valued in $\mathcal{M}_+$, the set  of nonnegative  measures on $\R^d$. 
In view of Equation \eqref{eq: return_wealth}, Proposition  \ref{lem: prop_p_bar}, and Proposition \ref{lem: unnorm_cond_law}, we then introduce the dynamic value function associated to 
Problem \eqref{eq: problem_U change} as
\beqs
v_k(x,z,\mu) & = &  \sup_{\al \in \Aq{k}(x,z)}  J_k(x,z,\mu,\al), \quad k \in [\![0,N]\!], \; (x,z) \in \Sc^q, \; \mu  \in  \mathcal{M}_+,
\enqs
with 
\beqs
J_k(x,z,\mu,\al) &=&   \overline \E \Big[ U \big(X^{k,x,\al}_N \big)  \pi_N^{k,\mu}(\R^d) \Big], \nonumber 
 \enqs
 where $X^{k,x,\al}$ is the solution to Equation \eqref{eq: return_wealth} on $[\![k,N]\!]$, starting at  $X^{k,x,\al}_k = x$ at time $k$,  controlled by $\al \in \Aq{k}(x,z)$, and 
 $(\pi_\ell^{k,\mu})_{\ell=k,\ldots,N}$ is the solution to \eqref{recurpi} on $\mathcal{M}_+$, starting from $\pi_k^{k,\mu}$ $=$ $\mu$, so that $V_0$ $=$ $v_0(x_0,x_0,\mu_0)$. 
 Here, $\Aq{k}(x,z)$ is the set of admissible investment strategies embedding the drawdown constraint: 
  $X_\ell^{k,x,\al}$ $\geq$ $qZ_\ell^{k,x,z,\al}$, $\ell$ $=$ $k,\ldots,N$, where the maximum wealth process 
  $Z^{k,x,z,\al}$ follows the dynamics: $Z_{\ell+1}^{k,x,z,\al}$ $=$ $\max[Z_\ell^{k,x,z,\al},X_{\ell+1}^{k,x,\al}]$, 
 $\ell$ $=$ $k,\ldots,N-1$, starting from $Z_k^{k,x,z,\al}$ $=$ $z$ at time $k$.  The dependence of the value function upon the 
 unnormalized filter $\mu$ means that the probability distribution on the drift is updated at each time step from  Bayesian learning by observing assets price.  
  
The dynamic programming equation associated to \eqref{eq: problem_U change} is then written in backward induction as

\begin{equation*}
\left\{
\begin{array}{ccl}
%\beqs \label{eq: formal_dyn_prog_gen}
v_N(x,z,\mu) & =&  U(x)  \mu(\R^d),   \\
v_k(x,z,\mu) &=&  \Sup_{\al \in \Aq{k}(x,z)}  \overline\E\Big[ v_{k+1} \big( X_{k+1}^{k,x,\al}, Z_{k+1}^{k,x,z,\alpha}, \pi_{k+1}^{k,\mu} \big)  \Big], \quad k=0,\ldots,N-1.
%\enqs
\end{array}
\right.
\end{equation*}
Recalling  Proposition \ref{lem: unnorm_cond_law} and Lemma \ref{lem: cond_a}, this dynamic programming system is written more explicitly as 
\begin{equation} \label{eq: dyn_prog_gen} 
\left\{
\begin{array}{ccl}
v_N(x,z,\mu) &=& U(x)  \mu(\R^d), \quad (x,z) \in \Sc^q, \; \mu  \in  \mathcal{M}_+,\\
v_k(x,z,\mu) &=& \Sup_{a \in A^q(x,z)}  \overline\E\Big[ v_{k+1} \Big( x\big(1 +  a^\prime \big(e^{R_{k+1}}- \mathbbm{1}_d \big)\big),\\
&& \hspace{5mm} \max\big[z,   x\big(1 +  a^\prime \big(e^{R_{k+1}}- \mathbbm{1}_d \big)\big) \big], \bar g(R_{k+1}-\cdot) \mu \Big)  \Big], 
\end{array}
\right.
\end{equation}
for $k$ $=$ $0,\ldots,N-1$.  Notice from  Proposition  \ref{lem: prop_p_bar} that the expectation in the above formula is only taken with respect to the noise $R_{k+1}$, which is distributed under $\overline{\P}$ 
according to the probability 
distribution $\nu$ with density $g$ on $\R^d$. 
 
\subsection{Special case: CRRA utility function} \label{subsec: CRRA}

In the case where the utility function is of  CRRA (Constant Relative Risk Aversion) type, i.e., 
\beq \label{UCRRA}
U(x) &=& \frac{x^p}{p}, \quad  x > 0, \; \mbox{ for some } 0 < p < 1, 
\enq
one can reduce the dimensionality of the problem. For this purpose, we introduce the process $\rho$ $=$ $(\rho_k)_k$  defined as the ratio of the wealth  over its maximum up to current  as:
\beqs \label{eq: rho_k+1_dyn}
\rho_{k}^{\alpha} &= &  \frac{\X{k}}{\Z{k}}, \quad  k=0,\ldots,N.
\enqs
This ratio process lies in the interval $[q,1]$ due to the maximum drawdown constraint. Moreover, recalling \eqref{eq: return_wealth}, and observing 
that $Z_{k+1}^\alpha$ $=$ $\max[Z_k^\alpha,X_{k+1}^\alpha]$, together with the fact that  $\frac{1}{\max[z,x]}$ $=$ $\min[\frac{1}{z},\frac{1}{x}]$, 
we notice that the ratio process $\rho$  can be written in inductive form as
\beqs
\rho_{k+1}^\alpha &=& \min\Big[1, \rho_k^\alpha\big(1 +  \alpha_k^\prime \big(e^{R_{k+1}}- \mathbbm{1}_d \big)\big) \Big], \quad k = 0,\ldots,N-1. 
\enqs

The following result states that the value function inherits the homogeneity property of the utility function.

\begin{lemma} \label{lem: homo_value_function}
For a utility function $U$ as in \eqref{UCRRA}, we have for all  $(x,z) \in \Sc^q$,  $\mu \in  \mathcal{M}_+$,   $k \in [\![0,N]\!]$,
\beqs
v_k(\lambda x,\lambda z,\mu) &=&  \lambda^{p} v_k(x,z,\mu), \quad \lambda > 0. 
\enqs
\end{lemma}
\noindent {\bf Proof.}
See Appendix \ref{Pr: lem: homo_value_function}.
\ep
 
In view of the above Lemma, we consider  the sequence of functions $w_k$, $k \in [\![0,N]\!]$, defined by 
 \beqs
 w_k(r,\mu) &=& v_k(r,1,\mu), \quad r \in [q,1],  \; \mu \in  \mathcal{M}_+, 
 \enqs
 so that $v_k(x,z,\mu)$ $=$ $z^p w_k(\frac{x}{z},\mu)$, and we call $w_k$ the reduced value function.  From the dynamic programming system satisfied by $v_k$, we immediately obtain the 
 backward system for $(w_k)_k$ as
 \begin{equation} \label{eq: dyn_prog_CRRA} 
\left\{
\begin{array}{ccl}
%\beqs \label{eq: formal_dyn_prog_gen}
w_N(r,\mu) & =&  \frac{r^p}{p}  \mu(\R^d), \quad r \in [q,1], \; \mu  \in  \mathcal{M}_+,      \\
w_k(r,\mu) &=&  \Sup_{a \in A^q(r)}  \overline\E\Big[ w_{k+1} \big(    
\min\big[1,   r\big(1 +  a^\prime \big(e^{R_{k+1}}- \mathbbm{1}_d \big)\big) \big], \bar g(R_{k+1}-\cdot) \mu \big)  \Big], 
%\enqs
\end{array}
\right.
\end{equation}
for $k$ $=$ $0,\ldots,N-1$, where   
\beqs
A^q(r) &=& \Big\{ a \in \Rd_+ : a^\prime \mathbbm{1}_d \leq 1- \frac{q}{r} \Big\}. 
\enqs
  
We end this section  by stating some properties on the reduced value function. 
 
\begin{lemma} \label{lem: v_prop}
For  any  $k \in [\![ 0, N ]\!]$, the reduced  value function $w_k$ is nondecreasing  and concave in $r$ $\in$ $[q,1]$.
\end{lemma}
\noindent {\bf Proof.}
See proof in Appendix \ref{Pr: lem: v_prop}.
\ep

\section{The Gaussian case} \label{sec: Gauss_case}

We consider in this section the Gaussian framework where the noise and the prior belief on the drift  are modeled according to a Gaussian distribution. 
In this special case, the Bayesian filtering is simplified into the Kalman filtering, and  the 
dynamic programming system is reduced to a finite-dimensional problem that will be solved numerically. It is convenient to deal directly with  the posterior distribution of the drift, i.e. the conditional law of the drift $B$ given the assets price observation, also called normalized filter. 
From \eqref{defmuk} and  Proposition \ref{lem: unnorm_cond_law}, it is given by the inductive relation
\beq \label{eq: rec_mu}
\mu_{k}(db) & = &   \frac{g(R_{k}-b) \mu_{k-1}(db)}{\int_{\Rd}g(R_{k}-b) \mu_{k-1}(db)}, \quad k=1,\ldots,N. 
\enq 
 
\subsection{Bayesian Kalman filtering} \label{subsec: Bay_filt_Gauss}
We assume that the probability law $\n$ of the noise  $\e{k}$ is  Gaussian: $\mathcal{N}(0,\covE)$, and so with density function
\beq \label{eq: fun_g}
g(r) &=& (2 \pi)^{-\frac{d}{2}}|\covE|^{-\frac{1}{2}}e^{-\frac{1}{2}r^\prime \covE^{-1} r}, \quad r \in \R^d. 
\enq
Assuming also that the prior distribution $\mu_0$ on the drift $B$ is Gaussian with mean $b_0$, and invertible covariance matrix $\covB$, we deduce by induction from 
\eqref{eq: rec_mu} that the posterior distribution $\mu_k$ is also Gaussian: $\mu_k$ $\sim$ ${\cal N}(\hat B_k,\Sigma_k)$, where $\hat B_k$ $=$ $\E[B|\F{k}^o]$ and $\Sigma_k$ satisfy the 
well-known  inductive relations:
\beq
\hat B_{k+1} &=& \hat B_k + K_{k+1} (R_{k+1} - \hat B_k), \quad k=0,\ldots,N-1 \label{estimB} \\
\Sigma_{k+1} &=& \Sigma_k  - \Sigma_k(\Sigma_k + \covE)^{-1}\Sigma_k,   \label{estimSigma}
\enq
where $K_{k+1}$ is the so-called Kalman gain given by 
\beq \label{eq: Kal_gain}
K_{k+1} &=& \Sigma_k(\Sigma_k + \covE)^{-1}, \quad k=0,\ldots,N-1.  
\enq
We have the initialization $\hat B_0$ $=$ $b_0$, and the notation for $\Sigma_k$ is coherent 
at time $k$ $=$ $0$ as it corresponds to the covariance matrice of $B$.  While the Bayesian estimation $\hat B_k$ of $B$ is updated from the current observation of the log-return $R_k$, notice that 
$\Sigma_k$ (as well as $K_k$) is deterministic, and is then equal to the covariance matrix of the error between $B$ and its Bayesian estimation, i.e. 
$\Sigma_k$ $=$ $\E[(B-\hat B_k)(B-\hat B_k)^\prime]$. Actually, we can explicitly compute $\Sigma_k$ by noting from Equation \eqref{eq: rec_mu} with $g$ as in \eqref{eq: fun_g} and 
$\mu_0 \sim \mathcal{N}(b_0, \covB)$ that
\beqs
	\mu_k \sim \frac{e^{-\frac{1}{2}\left(b-\left(\covB^{-1} + \covE^{-1} k  \right)^{-1}\left(\covE^{-1}\sum_{j=1}^{k}R_j+\covB^{-1}b_0  \right)  \right)\left(\covB^{-1} + \covE^{-1} k  \right) \left(b-\left(\covB^{-1} + \covE^{-1} k  \right)^{-1}\left(\covE^{-1}\sum_{j=1}^{k}R_j+\covB^{-1}b_0  \right) \right) }}{(2\pi)^{\frac{d}{2}}|(\covB^{-1}+\covE^{-1}k)^{-1}|^{\frac{1}{2}}}.
\enqs
By identification, we then get  
\beq \label{eq: expl_Sigmak}
\Sigma_k = (\covB^{-1}+ \ivCovE k)^{-1} = \covB (\covE + \covB k)^{-1} \covE.
\enq

Moreover, the innovation process $(\tilde\epsilon_k)_k$, defined as 
\beq \label{innov} 
\te{k+1} &=& R_{k+1}- \E[R_{k+1} | \F{k}^o] \; = \; R_{k+1} -  \hB{k}, \quad k=0,\ldots,N-1,
\enq
is  a $\Fb^o$-adapted Gaussian process. Each $\tilde\epsilon_{k+1}$ is independent of $\F{k}^0$ (hence  $\tilde\epsilon_k$, $k$ $=$ $1,\ldots,N$ are mutually independent), and is a centered Gaussian vector  with covariance matrix:
\beqs
\tilde\epsilon_{k+1} & \sim &  {\cal N}\big(0,\tilde\covE_{k+1}\big), \quad \mbox{ with } \; \tilde\covE_{k+1} \; = \;  \Sigma_k + \covE.   
\enqs
We refer to \cite{Kalman1960} and \cite{Kalman1961} for these classical properties about the Kalman filtering and the innovation process.

\begin{remark} \label{remBk}
{\rm From \eqref{estimB}, and \eqref{innov}, we see that the Bayesian estimator $\hat B_k$ follows the dynamics
\beqsalac
\hat B_{k+1} &= \; \hat B_k + K_{k+1} \tilde\epsilon_{k+1},\;\;\;\; k=0,\ldots,N-1\\
\hat B_0 &= b_0, 
\enqsalac
which implies in particular that $\hat B_k$ has a Gaussian distribution with mean $b_0$, and covariance matrix satisfying 
\begin{align*}
\mbox{Var}(\hat B_{k+1}) & = \; \mbox{Var}(\hat B_{k}) + K_{k+1}(\Sigma_k + \Gamma) K_{k+1}^\prime \: = \; \mbox{Var}(\hat B_{k})  + \Sigma_k(\Sigma_k + \Gamma)^{-1}\Sigma_k. 
\end{align*}
Recalling the inductive relation \eqref{estimSigma} on $\Sigma_k$, this shows that $\mbox{Var}(\hat B_{k})$ $=$ $\Sigma_0 - \Sigma_k$.  
Note that, from Equation \eqref{estimSigma}, $(\Sigma_k)_k$ is a decreasing sequence which ensures that $\mbox{Var}(\hat B_{k})$ is positive semi-definite and is nondecreasing with 
time $k$. 
}
\epR
\end{remark}

\subsection{Finite-dimensional dynamic programming equation} \label{subsec: Der_dyn_prog_eq}
From  \eqref{innov}, we see that our initial portfolio selection Problem \eqref{eq: pb_init} can be reformulated as a full observation problem with state dynamics given by 
\begin{equation} \label{dynXB} 
\left\{
\begin{array}{ccl}
\X{k+1} &=&  \X{k} \Big( 1 + \al_k^\prime \big(e^{\hat B_k + \tilde\epsilon_{k+1}}- \mathbbm{1}_d \big) \Big),   \\
\hat B_{k+1} &=& \hat B_k + K_{k+1} \tilde\epsilon_{k+1},  \quad  k =0,\ldots,N-1.  
\end{array}
\right.
\end{equation}

We then define the  value function on $[\![0,N]\!]\times\Sc^q\times\R^d$ by 
\beqs
\tilde v_k(x,z,b) &=&   \Sup_{\al \in \Aq{k}(x,z)} \E \big[ U(X_N^{k,x,b,\alpha}) \big], \quad k \in [\![0,N]\!], \;  (x,z) \in \Sc^q, \; b \in \R^d,
\enqs
where the pair $(X^{k,x,b,\alpha},\hat B^{k,b})$ is the process solution to \eqref{dynXB} on  $[\![k,N]\!]$, starting from $(x,b)$ at time $k$, so that $V_0$ $=$ $\tilde v_0(x_0,x_0,b_0)$.  
The associated dynamic programming system satisfied by the sequence $(\tilde v_k)_k$ is 
\begin{numcases}{}  
   \tilde v_N(x,z,b) \; = \;  U(x),  \quad (x,z) \in \Sc^q, \; b \in \R^d, \hspace{11.5cm} \nonumber \\
   \tilde v_k(x,z,b) \;\; = \; \Sup_{a \in A^q(x,z)}  \E\Big[ \tilde v_{k+1} \Big( x\big(1 +  a^\prime \big(e^{ b + \tilde\epsilon_{k+1}}- \mathbbm{1}_d \big)\big), \nonumber \\  
\hspace{5cm} \max\big[z,   x\big(1 +  a^\prime \big(e^{b + \tilde\epsilon_{k+1}}- \mathbbm{1}_d \big)\big) \big], b + K_{k+1}\tilde\epsilon_{k+1} \Big) \Big], \nonumber 
\end{numcases}
for $k$ $=$ $0,\ldots,N-1$. Notice that in the above formula, the expectation is taken with respect to the innovation vector $\tilde\epsilon_{k+1}$, which is distributed according to 
${\cal N}(0,\tilde\covE_{k+1})$. 

Moreover, in the case of CRRA utility functions $U(x)$ $=$ $x^p/p$, and similarly as in Section \ref{subsec: CRRA}, we have the dimension reduction with
\beqs
\tilde w_k(r,b) &=& \tilde v_k(r,1,b), \quad r \in [q,1],  \; b \in  \R^d,  
 \enqs
 so that $\tilde v_k(x,z,b)$ $=$ $z^p \tilde w_k(\frac{x}{z},b)$, and this reduced value function satisfies the backward system on $[q,1]\times\R^d$: 
 \begin{equation*} 
\left\{
\begin{array}{ccl}
%\beqs \label{eq: formal_dyn_prog_gen}
\tilde w_N(r,b) & =&  \frac{r^p}{p}, \quad r \in [q,1], \;  b   \in  \R^d,      \\
\tilde w_k(r,b) &=&  \Sup_{a \in A^q(r)}  \E\Big[ \tilde w_{k+1} \Big(    
\min\big[1,   r\big(1 +  a^\prime \big(e^{b + \tilde\epsilon_{k+1}}- \mathbbm{1}_d \big)\big) \big], b + K_{k+1}\tilde\epsilon_{k+1} \Big)  \Big], 
%\enqs
\end{array}
\right.
\end{equation*}
for $k$ $=$ $0,\ldots,N-1$. 
 
\begin{remark}[No short-sale constrained Merton problem] \label{remMerton} 
{\rm In the limiting case when $q$ $=$ $0$, the drawdown constraint is reduced  to a  non-negativity constraint on the wealth process, and by Lemma \ref{lem: cond_a}, this means 
a no-short selling and no borrowing constraint on the portfolio strategies. When the drift $B$ is also known, equal to $b_0$, and for a CRRA utility function,  let us then consider the 
corresponding constrained Merton problem with value function denoted by $v_k^M$, $k$ $=$ $0,\ldots,N$, which  satisfies the standard backward recursion from dynamic programming:  
\beqalac \label{eq: Merton_pb}
		v_N^M(x) & = \frac{x^p}{p}, \;\;x>0, \\
		v_k^M(x) &= \Sups{a^\prime \mathbbm{1}_d \leq 1}{a \in [0,1]^d}  \E\Big[ v_{k+1}^M \big( x\big(1 +  a^\prime \big(e^{ b_0 + \epsilon_{k+1}}- \mathbbm{1}_d \big)\big)\Big], 
		\quad k=0,\ldots,N-1. 
\enqalac 
Searching for a solution of the form $v_k^M(x)$ $=$ $K_k x^p/p$, with $K_k \geq 0$ for all $k \in [\![0, N]\!]$, we see that the sequence $(K_k)_k$ satisfies the recursive relation:
\beqs
	K_{k} &=&  S K_{k+1}, \quad k=0,\ldots,N-1, 
\enqs
starting from $K_N$ $=$ $1$, where   
\beqs
S  &:=&  \Sups{a^\prime \mathbbm{1}_d \leq 1}{a \in [0,1]^d}  \E\Big[ \Big( 1 + a^\prime\big(e^{b_0 + \epsilon_{1}}-\mathbbm{1}_d \big) \Big)^p\Big],
\enqs
by recalling that $\epsilon_{1}$, $\dots$, $\epsilon_{N}$ are i.i.d. random variables. It follows that the value function of the constrained Merton problem, unique solution to the dynamic programming system  \eqref{eq: Merton_pb}, is equal to 
\beqs
v_k^M(x) &=& S^{N-k} \frac{x^p}{p}, \quad k=0,\ldots,N,  
\enqs
and the constant optimal control is given by
\beqs \label{eq: Merton controls}
a^M_k &=&  \displaystyle\argmax_{\substack{a^\prime \mathbbm{1} \leq 1 \\ a \in [0,1]^d}}  \E\left[ \left ( 1 + a^\prime \left(e^{R_1}-\mathbbm{1}_d\right)   \right) ^p\right]\;\; 
k=0, \dots, N-1.
\enqs
}
\epR
\end{remark} 
 
\section{Deep learning numerical resolution} \label{sec: Deep learning numerical resolution}
In this section, we exhibit numerical results  to promote the benefits of learning from new information. To this end, we compare the learning strategy (Learning) to the non-learning one (Non-Learning) in the case of the CRRA utility function and the Gaussian distribution for the noise. The prior probability distribution of $B$ is the Gaussian distribution $\mathcal{N}(b_0, \covB)$ for Learning while it is the Dirac distribution concentrated at $b_0$ for Non-Learning.

We use deep neural network techniques to compute numerically the optimal solutions for both Learning and Non-Learning. To broaden the analysis, in addition to the learning and non-learning strategies, we have computed an ''admissible" equally weighted (EW) strategy. More precisely, this EW strategy will share the quantity $X_k-qZ_k$ equally among the $d$ assets. Eventually, we show numerical evidence that the Non-Learning converges to the optimal strategy of the constrained Merton problem, when the loss aversion parameter $q$ vanishes.

\subsection{Architectures of the deep neural networks} \label{subsec: arch_dnn}
Neural networks (NN) are able to approximate nonlinear continuous functions, typically the value function and controls of our problem. The principle is to use a large amount of data to train the NN so that it progressively comes close to the target function. It is an iterative process in which the NN is tuned on a training set, then tested on a validation set to avoid over-fitting. For more details, see for instance \cite{hornik1991} and \cite{Geron2019}. 

\begin{figure}[H]
   \begin{minipage}{0.48\textwidth}
     \centering
     \includegraphics[width=\linewidth]{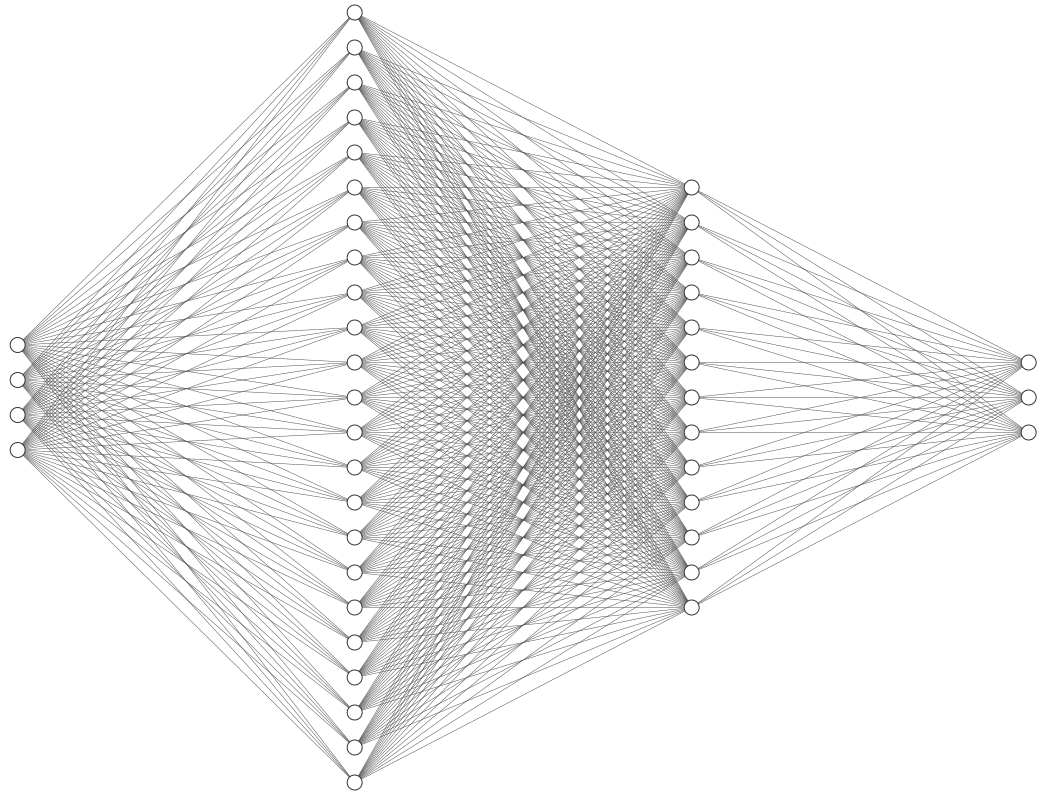}
     \caption{$A_{NN}$ architecture with $d=3$ assets}\label{Fig:NN_Control}
   \end{minipage}\hfill
   \begin{minipage}{0.48\textwidth}
     \centering
     \includegraphics[width=\linewidth]{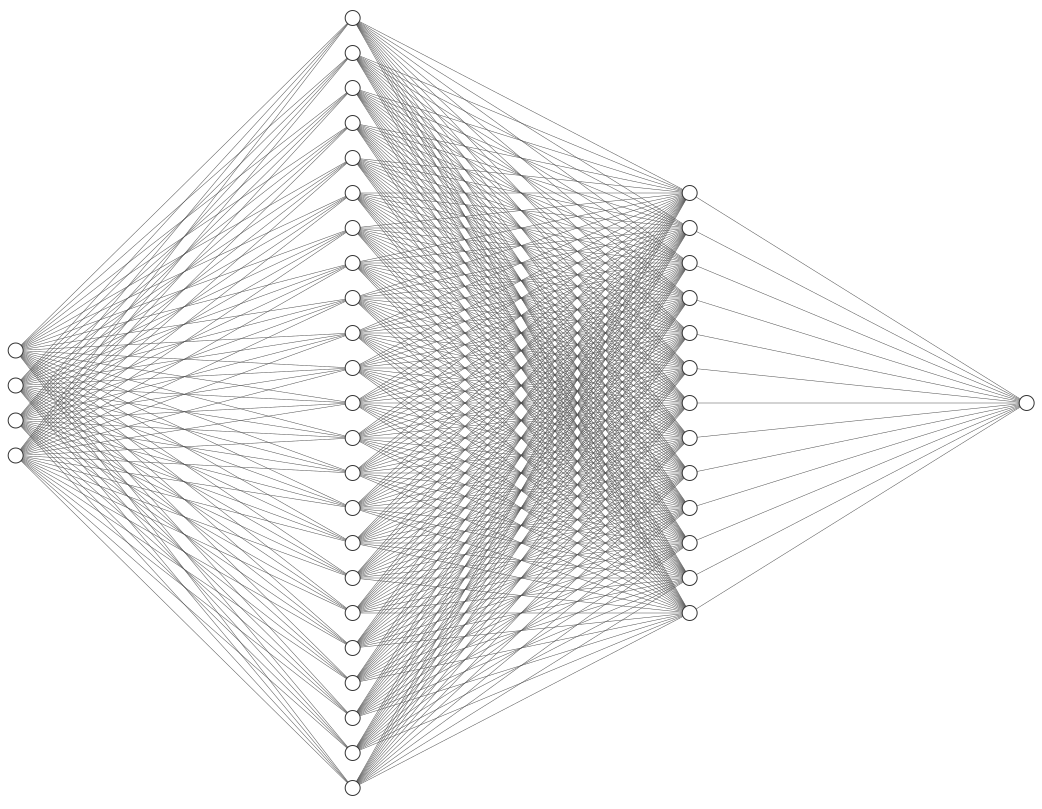}
     \caption{$VF_{NN}$ architecture with $d=3$ assets}\label{Fig:NN_VF}
   \end{minipage}
\end{figure}

The algorithm we use, relies on two dense neural networks: the first one is dedicated to the controls ($A_{NN}$) and the second one to the value function ($VF_{NN}$). Each NN is composed of four layers: an input layer, two hidden layers and an output layer: 
\begin{itemize}
\item[(i)] The input layer is $d+1$-dimensional since it embeds the conditional expectations of each of the $d$ assets and the ratio of the current wealth to the current historical maximum $\rho$. 
\item[(ii)] The two hidden layers give the NN the flexibility to adjust its weights and biases to approximate the solution. From numerical experiments, we see that, given the complexity of our problem, a first hidden layer with $d+20$ neurons and a second one with $d+10$ are a good compromise between speed and accuracy. 
\item[(iii)] 
The output layer is $d$-dimensional for the controls, one for each asset representing the weight of the instrument, and is one-dimensional for the value function. See Figures \ref{Fig:NN_Control} and \ref{Fig:NN_VF} for an overview of the NN architectures in the case of $d=3$ assets.
\end{itemize}

We follow the indications in \cite{Geron2019} to setup and define the values of the various inputs of the neural networks which are listed  in Table \ref{tab: NN parameters}.

\begin{table}[!t]
\makebox[1 \textwidth][c]{       %centering table
\resizebox{1. \textwidth}{!}{
      \begin{tabular}{lcc}       
		\hline\noalign{\smallskip}
         Parameter & $\mathbf{A_{NN}}$ & $\mathbf{VF_{NN}}$ \\
		\noalign{\smallskip}\hline\noalign{\smallskip}
         	Initializer & uniform$(0, 1)$ & He\_uniform \\		
        	Regularizers & L2 norm & L2 norm\\       
         	Activation functions & Elu and Sigmoid for output layer & Elu and Sigmoid for output layer \\         	
         	Optimizer & Adam & Adam \\
         	Learning rates: step N-1 & 5e-3 & 1e-3 \\        	
         	\hspace{1.8cm} steps k = 0,...,N-2 & 6.25e-4 & 5e-4 \\         	
        	Scale & 1e-3 & 1e-3 \\       	
        	Number of elements in a training batch & 3e2 & 3e2  \\     
         	Number of training batches & 1e2 & 1e2  \\         	
        	Size of the validation batches & 1e3 & 1e3 \\         			       	        
         	Penalty constant & 3e-1 & NA \\         	
         	Number of epochs: step N-1 & 2e3 & 2e3 \\        	
         	\hspace{2.3cm} steps k = 0,...,N-2 & 5e2 & 5e2 \\        	
         	Size of the training set: step N-1 & 6e7 & 6e7  \\			
			\hspace{2.8cm} steps k = 0,...,N-2 & 1.5e7 & 1.5e7 \\						
         	Size of the validation set: step N-1 & 2e6 & 2e6  \\        	
         	\hspace{3.05cm} steps k = 0,..., N-2 & 5e5 & 5e5 \\	            	
        \noalign{\smallskip}\hline\noalign{\smallskip}
      \end{tabular}   
      }
      }
      \caption{Parameters for the neural networks of the controls $\mathbf{A_{NN}}$ and the value function $\mathbf{VF_{NN}}$.}
      \label{tab: NN parameters}
\end{table}

To train the NN, we simulate the input data. For the conditional expectation $\hB{k}$, we use its time-dependent Gaussian distribution (see Remark \ref{remBk}):  
$\hat B_k$ $\sim$ $\mathcal{N}(b_0,\Sigma_0 - \Sigma_k)$, with $\Sigma_k$ as in Equation \eqref{eq: expl_Sigmak}. 
On the other hand,  the training of $\rho$ is drawn from the uniform distribution between $q$ and $1$, the interval where it lies according to the maximum drawdown constraint. 

\subsection{\emph{Hybrid-Now} algorithm}
We use the \emph{Hybrid-Now} algorithm developped in \cite{Bachouch2018b} in order to solve numerically our problem. This algorithm combines optimal policy estimation by neural networks and dynamic programming principle which suits the approach we have developped in Section \ref{sec: Gauss_case}.

With the same notations as in Algorithm \ref{al: Hybrid} detailed in the next insert, at time $k$, the algorithm computes the proxy of the optimal control $\hat{\alpha}_{k}$ with $A_{NN}$, using the known function $\hat{V}_ {k+1}$ calculated the step before, and uses $V_{NN}$ to obtain a proxy of the value function $\hat{V}_{k}$. Starting from the known function $\hat{V}_N:=U$ at terminal time $N$, the algorithm computes sequentially $\hat{\alpha}_k$ and $\hat{V}_k$ with backward iteration until time $0$. This way, the algorithm loops to build the optimal controls and the value function pointwise and gives as output the optimal strategy, namely the optimal controls from $0$ to $N-1$ and the value function at each of the $N$ time steps.

The maximum drawdown constraint is a time-dependent constraint on the maximal proportion of wealth to invest (recall Lemma \ref{lem: cond_a}). 
In practice, it is a constraint on the sum of weights of each asset or 
equivalently on the output of $A_{NN}$. For that reason, we have implemented an appropriate penalty function that will reject undesirable values: 
\beqs
	G_{Penalty}(A, r) =   K {\rm max}\left( |A|_{_1} \leq 1-\frac{q}{r}, 0 \right), \;\; A \in [0,1]^d, \;\; r \in [q,1]. 
\enqs
This penalty function ensures that the strategy respects the maximum drawdown constraint at each time step, when the parameter $K$ is chosen sufficiently large. 
 
\begin{algorithm}[H] \label{al: Hybrid}
\caption{\emph{Hybrid-Now}}
\DontPrintSemicolon
 \KwInput{the training distributions $\mu_{Unif}$ and $\mu_{Gauss}^k$; \\ 
 \Comment*[r]{$\mu_{Unif} = \mathcal{U}(q,1)$ \\ $\mu_{Gauss}^k =  \mathcal{N} \left(b_0, \covB-\Sigma_k\right)$}}
 \KwOutput{\\
- estimate of the optimal strategy $\left(\hat{a}_k\right)_{k=0}^{N-1}$;\\
- estimate of the value function $\left(\hat{V}_k\right)_{k=0}^{N-1}$;}
Set $\hat{V}_N = U$;

\For{ $k = N-1, \dots , 0$}
{Compute:
\beqs
	\hat{\beta}_k \in \underset{\beta \in \R^{2d^2+56d+283}}{ \text{argmin}} \E \left[ G_{Penalty}(A_{NN}(\rho_k, \hB{k}; \beta),\rho_k)- \hat{V}_{k+1} \left(\rho_{k+1}^{\beta}, \hB{k+1}  \right) \right]
\enqs
where $\rho_k \sim \mu_{Unif}$, $\hB{k} \sim \mu_{Gauss}^k$, \\ $\hat{B}_{k+1} = \tilde{H}_k(\hat{B}_k, \te{k+1})$ and $\rho_{k+1}^{\beta} = F \left( \rho_k, \hB{k}, A_{NN}\left( \rho_k, \hB{k}; \beta \right), \te{k+1} \right)$;\\

\Comment*{$F(\rho,b,a,\epsilon) = {\rm min} \left( 1, \rho \left( 1 + \sum_{i=1}^{d} a^i \left(e^{b^i + \epsilon^i}-1 \right) \right)\right)$ \\ $\tilde{H}_k(b,\epsilon) = b + \covB(\covE+\covB k)^{-1}\epsilon$}

Set $\hat{a}_k = A_{NN} \left(.; \hat{\beta}_k \right);$ \\
\Comment*[r]{$\hat{a}_k$ is the estimate of the optimal control at time $k$.}
Compute:
\beqs
	\hat{\theta}_k \in \underset{\theta \in \R^{2d^2+54d+261}}{ \text{argmin}} \E \left[ \left(\hat{V}_{k+1} \left(\rho^{\hat{\beta}_k}_{k+1}, \hB{k+1}\right) - VF_{NN} \left(\rho_k, \hat{B}_k ; \theta \right) \right)^2\right]
\enqs

Set $\hat{V}_k = VF_{NN} \left(., \hat{\theta}_k \right)$ ; \\
\Comment*[r]{$\hat{V}_k$ is the estimate of the value function at time $k$.} 
}
\end{algorithm}

A major argument  behind the choice of this algorithm is that, it is particularly relevant  for problems in which the neural network approximation of the controls and value function at time $k$, are close to the ones at time $k+1$. This is what we expect in our case. We can then take a small learning rate for the Adam optimizer which enforces the stability of the parameters' update during the gradient-descent based learning procedure.

\subsection{Numerical results} \label{subsec: num_res}
In this section, we explain the setup of the simulation and exhibit the main results. We have used Tensorflow 2 and deep learning techniques for Python developped in \cite{Geron2019}. We  consider $d=3$ risky assets and a riskless asset whose return is assumed to be $0$, on a 1-year investment horizon for the sake of simplicity. We consider  $24$ portfolio rebalancing during the 1-year period, i.e.,  one every two weeks. This means that we have $N$ $=$ $24$  steps in the training of our neural networks. The parameters used in the simulation are detailed in 
Table \ref{tab: simulation parameters}.

\begin{table}[!t]
      \centering 
      \scalebox{1.}{\begin{tabular}{lc} 
        \hline\noalign{\smallskip}
         Parameter & Value\\
		\noalign{\smallskip}\hline\noalign{\smallskip}
         	Number of risky assets $d$ &$3$  \\		
        	Investment horizon in years $T$ &$1$\\       
         	Number of steps/rebalancing $N$ &$24$ \\         	
         	Number of simulations/trajectories $\tilde{N}$ &$1000$ \\         	
         	Degree of the CRRA utility function $p$ &$0.8$ \\         	
         	Parameter of risk aversion $q$ &0.7  \\         	
         	Annualized expectation of the drift B &$\begin{bmatrix} 0.05 & 0.025 & 0.12 \end{bmatrix}$  \\        	
         	Annualized covariance matrix of the drift $B$ &$\begin{bmatrix} 0.2^2 & 0 & 0\\ 0 & 0.15^2 & 0\\ 0 & 0 & 0.1^2 \end{bmatrix}$ \\          	
         	Annualized volatility of $\e{}$ &$\begin{bmatrix} 0.08 & 0.04 & 0.22 \end{bmatrix}$  \\        	
         	Correlation matrix of $\e{}$ & $\begin{bmatrix} 1 & -0.1 & 0.2\\ -0.1 & 1 & -0.25\\ 0.2 & -0.25 & 1 \end{bmatrix}$  \\         	        	
         	Annualized covariance matrix of the noise $\e{}$ &$\begin{bmatrix} 0.0064 & -0.00032 & 0.00352\\ -0.00032 & 0.0016 & -0.0022\\ 0.00352 & -0.0022 & 0.0484 \end{bmatrix}$  \\ 
        \noalign{\smallskip}\hline\noalign{\smallskip}
      \end{tabular} }    
      \caption{Values of the parameters used in the simulation.}
      \label{tab: simulation parameters}
\end{table}

First, we show the numerical results for  the learning and the non-learning strategies by presenting a performance and an allocation analysis in Subsection \ref{subsec: L_and_NL_strategies}. Then, we add the admissible constrained EW to the two previous ones and use this neutral strategy as a benchmark in Subsection \ref{subsec: L_NL_and_constrained_EW_strategies}. Ultimately, in Subsection \ref{subsec: NL_and_Merton_strategies}, we illustrate numerically the convergence of the non-learning strategy to the constrained 
Merton problem when the loss aversion parameter $q$ vanishes.

\subsubsection{Learning and non-learning strategies} \label{subsec: L_and_NL_strategies}
\begin{figure}[H]
\makebox[1 \textwidth][c]{       %centering table
\resizebox{1. \textwidth}{!}{
	\begin{minipage}{0.49\textwidth}
	\centering
	\includegraphics[width=1.05\linewidth]{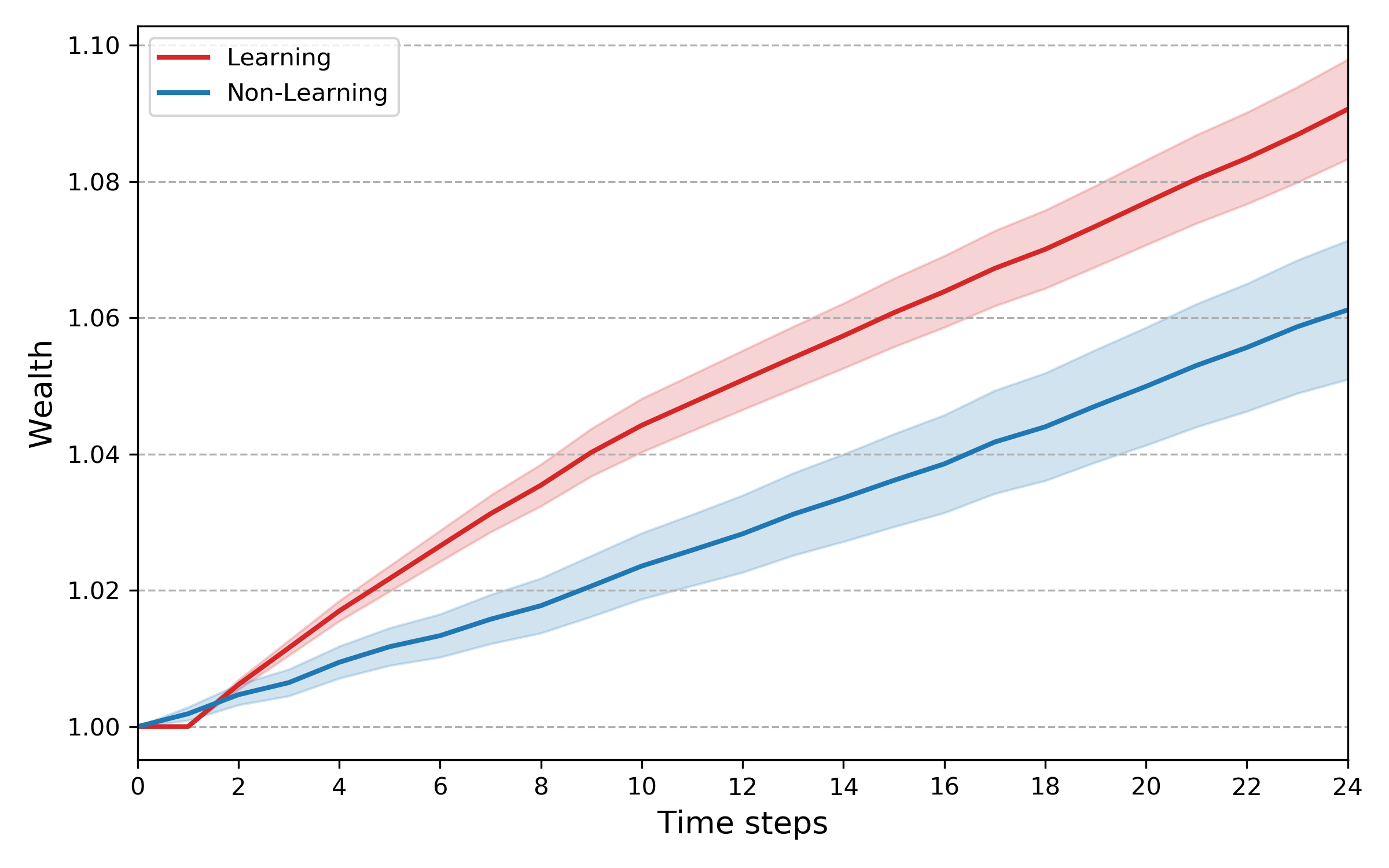}
	\caption{Historical Learning and Non-Learning levels with a 95\% confidence interval. }\label{Fig:MC_mean_std_L_NL}
	\end{minipage}\hspace{0.3cm}
   \begin{minipage}{0.49\textwidth}
     \centering
     \includegraphics[width=1.05\linewidth]{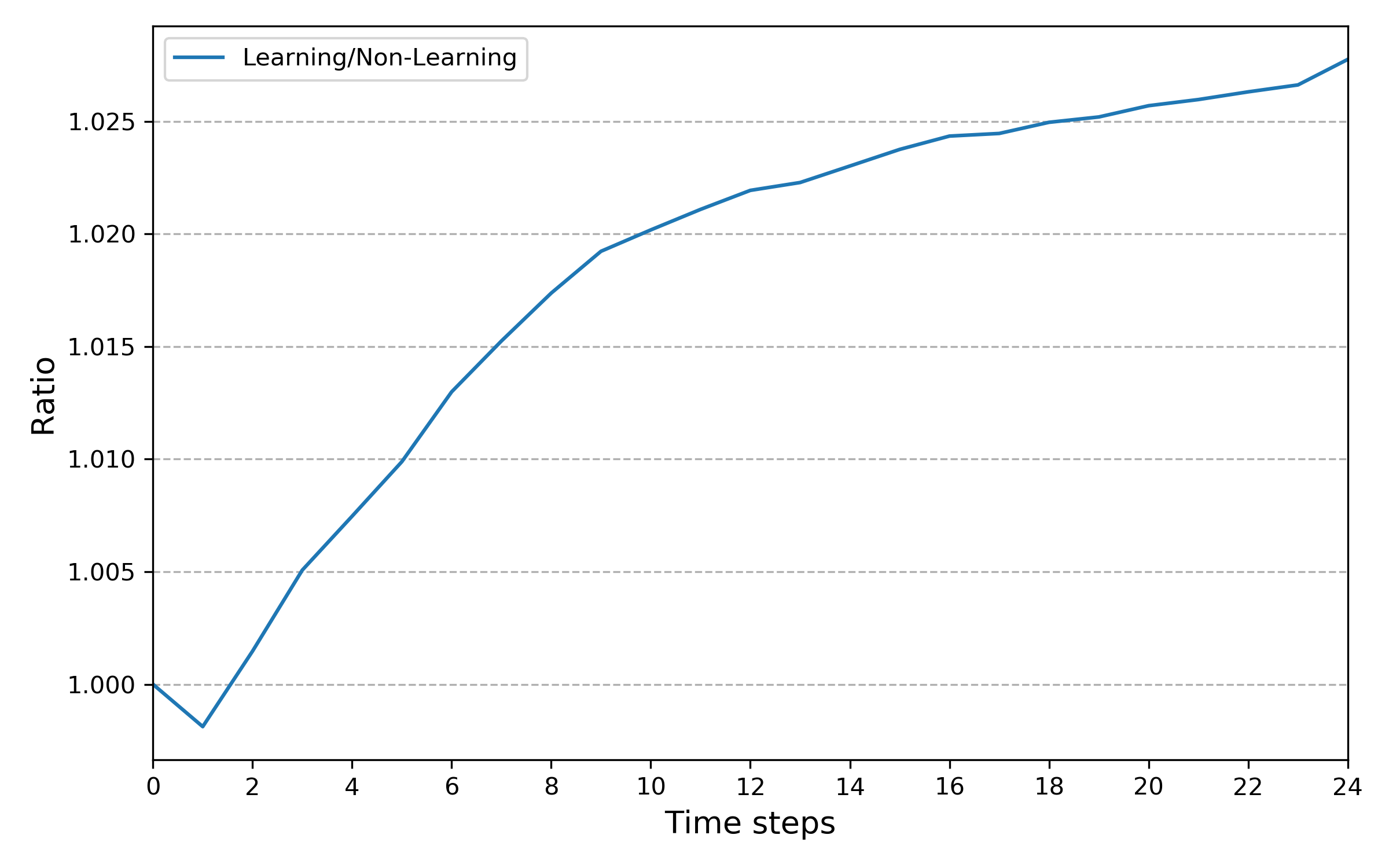}
     \caption{Historical ratio of Learning over Non-Learning levels. }\label{Fig:MC_Ratio_Learning-Non-learning}
     \end{minipage}
     }
     }
\end{figure}

We simulate $\tilde{N}=1000$ trajectories for each strategy and exhibit the performance results with an  initial wealth $x_0=1$. Figures \ref{Fig:MC_mean_std_L_NL} illustrates the average historical level of the learning and non-learning strategies with a $95\%$ confidence interval. Learning outperforms significantly Non-Learning with a narrower confidence interval revealing that less uncertainty surrounds Learning performance, thus yielding less risk.

An interesting phenomenon, visible in Fig. \ref{Fig:MC_mean_std_L_NL}, is the nearly flat curve for Learning between time $0$ and time $1$. Indeed, whereas Non-Learning starts investing immediately, Learning adopts a safer approach and needs a first time step before allocating a significant proportion of wealth. Given the level of uncertainty surrounding $b_0$, this first step allows Learning to fine-tune its allocation by updating the prior belief with the first return available at time $1$. On the contrary, Non-Learning, which cannot update its prior, starts investing at time $0$.

Fig. \ref{Fig:MC_Ratio_Learning-Non-learning} shows the ratio of Learning over Non-Learning. A ratio greater than one means that Learning outperforms Non-Learning and underperforms when less than one. It shows the significant outperformance of Learning over Non-Learning except during the first period where Learning was not significantly invested and Non-Learning had a positive return. Moreover, this graph reveals the typical increasing concave curve of the value of information described in \cite{Keppo2018}, in the context of investment decisions and costs of data analytics, and in \cite*{dfjnhp19BL} in the resolution of the Markowitz portfolio selection problem using a Bayesian learning approach.

Table \ref{tab: Statistics_L_NL} gathers relevant statistics for both Learning and Non-Learning such as: average total performance, standard deviation of the terminal wealth $X_T$, Sharpe ratio computed as average total performance over standard deviation of terminal wealth. The maximum drawdown (MD) is examined through two statistics: noting $MD^{\tilde{s}}_\ell$ the maximum drawdown of the $\ell$-$th$ trajectory of a strategy $\tilde{s}$, the average MD is defined as,
\beqs
	{\rm Avg\;MD}^{\tilde{s}} = \frac{1}{\tilde{N}} \sum_{\ell=1}^{\tilde{N}} {\rm MD}^{\tilde{s}}_\ell,
\enqs
for $\tilde{N}$ trajectories of the strategy $\tilde s$, and the worst MD is defined as, 
\beqs
	{\rm Worst\;MD}^{\tilde{s}} = {\rm min} \left( MD^{\tilde{s}}_1, \dots, MD^{\tilde{s}}_{\tilde{N}}  \right) .
\enqs
Finally, the Calmar ratio, computed as the ratio of the average total performance over the average maximum drawdown, is the last statistic exhibited. \\

With the simulated dataset, Learning delivered, on average, a total performance of $9.34\%$ while Non-Learning only $6.40\%$. Integrating the most recent information yielded a $2.94\%$ excess return. Moreover, risk metrics are significantly better for Learning than for Non-Learning. Learning exhibits a lower standard deviation of terminal wealth than Non-Learning ($11.88\%$ versus $16.67\%$), with a difference of $4.79\%$. More interestingly, the maximum drawdown is notably better controlled by Learning than by Non-Learning, on average ($-1.53\%$ versus $-6.54\%$) and in the worst case ($-11.74\%$ versus $-27.18\%$). This result suggests that learning from new observations, helps the strategy to better handle the dual objective of maximizing total wealth while controlling the maximum drawdown. We also note that learning improves the Sharpe ratio by $104.95\%$ and the Calmar ratio by $525.26\%$.

\begin{table}[!t]
      \centering 
      \scalebox{0.8}{\begin{tabular}{lccc} 
		\hline\noalign{\smallskip}
          Statistic & Learning & Non-Learning & Difference\\
         \noalign{\smallskip}\hline\noalign{\smallskip}        	
         	Avg total performance & 9.34\%  &  6.40\%  &  2.94\%  \\		
        	Std dev. of $X_T$     &11.88\%  & 16.67\%  & -4.79\%  \\      	
        	Sharpe ratio          & 0.79    &  0.38    &104.95\%  \\       
         	Avg MD     &-1.53\%  & -6.54\%  &  5.01\%  \\ 
         	Worst MD   &-11.74\%  &-27.18\%  & 15.44\%  \\      	
         	Calmar ratio          & 6.12    &  0.98    &525.26\%  \\			
        \noalign{\smallskip}\hline\noalign{\smallskip}
      \end{tabular}}    
      \caption{Performance metrics: Learning and Non-Learning. The difference for ratios are computed as relative improvement.} 
      \label{tab: Statistics_L_NL}
\end{table}

\begin{figure}[H]
\makebox[1.0 \textwidth][c]{       %centering table
\resizebox{1.0\textwidth}{!}{
     \includegraphics[width=1.05\linewidth]{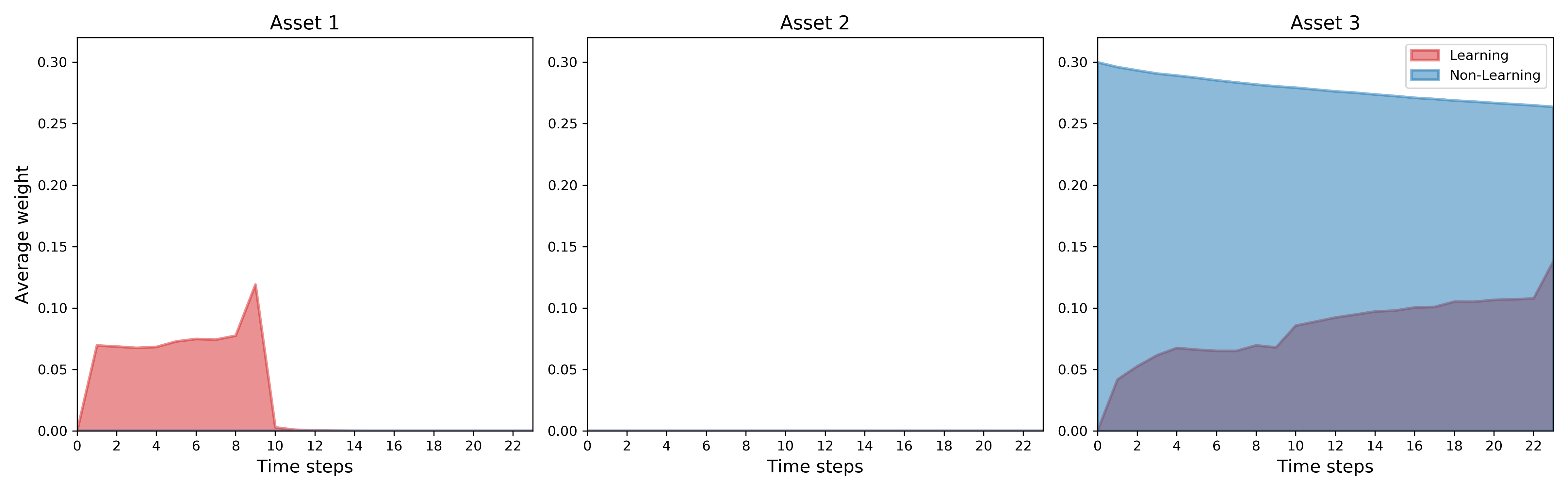}
     }
     }
     \caption{Historical Learning and Non-Learning asset allocations.}\label{Fig:Weights_Area_L_NL}
\end{figure}

\begin{figure}[H]
\makebox[1.0 \textwidth][c]{       %centering table
\resizebox{1.0 \textwidth}{!}{
     \includegraphics[width=1.05\linewidth]{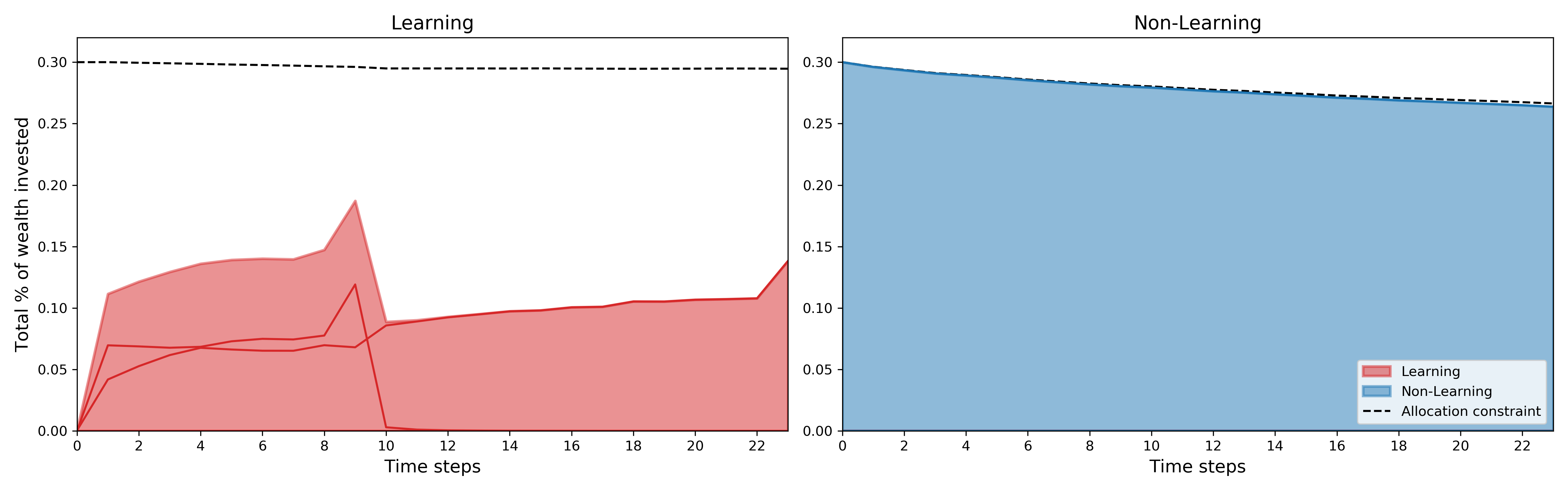}
     }
     }
     \caption{Historical Learning and Non-Learning total allocations.}\label{Fig:Weights_Sums_L_NL}
\end{figure}

Fig. \ref{Fig:Weights_Area_L_NL} and \ref{Fig:Weights_Sums_L_NL} focus more precisely on the portfolio allocation. The graphs of Fig. \ref{Fig:Weights_Area_L_NL} show the historical average allocation for each of the three risky assets.  First, none of the strategies invests in Asset 2 since it has the lowest expected return according to the prior, see Table \ref{tab: simulation parameters}. Whereas Non-Learning focuses on Asset 3, the one with the highest expected return, Learning performs an optimal allocation between Asset 1 and Asset 3 since this strategy is not stuck with the initial estimate given by the prior. Therefore, Learning invests little at time 0, then balances nearly equally both Assets 1 and 3, and then invests only in Asset 3 after time step 12. Instead, Non-Learning is investing only in Asset 3, from time $0$ until  the end of the investment horizon.

The curves in Fig. \ref{Fig:Weights_Sums_L_NL} recall each asset's optimal weight, but the main features are the colored areas that represent the average historical total percentage of wealth invested by each strategy. The dotted line represents the total allocation constraint they should satisfy to be admissible. To satisfy the maximum drawdown constraint, admissible strategies can only invest in risky assets the proportion of wealth that, in theory, could be totally lost. This explains why the non-learning strategy invests at full capacity on the asset that has the maximum expected return according to the prior distribution.

We clearly see that both strategies satisfy their respective constraints. Indeed, looking at the left panel, Learning is far from saturating the constraint. It has invested, on average, roughly $10\%$ of its wealth while its constraint was set around $30\%$. Non-learning invests at full capacity saturating its allocation constraint. Remark that this constraint is not a straight line since it depends on the value of the ratio: current wealth over current historical maximum, and evolves according to time.

\subsubsection{Learning, non-learning and constrained equally-weighted strategies} \label{subsec: L_NL_and_constrained_EW_strategies}
In this section, we add a simple constrained equally-weighted (EW) strategy to serve as a benchmark for both Learning and Non-Learning.
At each time step, the constrained EW strategy invests, equally across the three assets, the proportion of wealth above the threshold $q$.

\begin{figure}[H]
\centering
\includegraphics[scale=0.4]{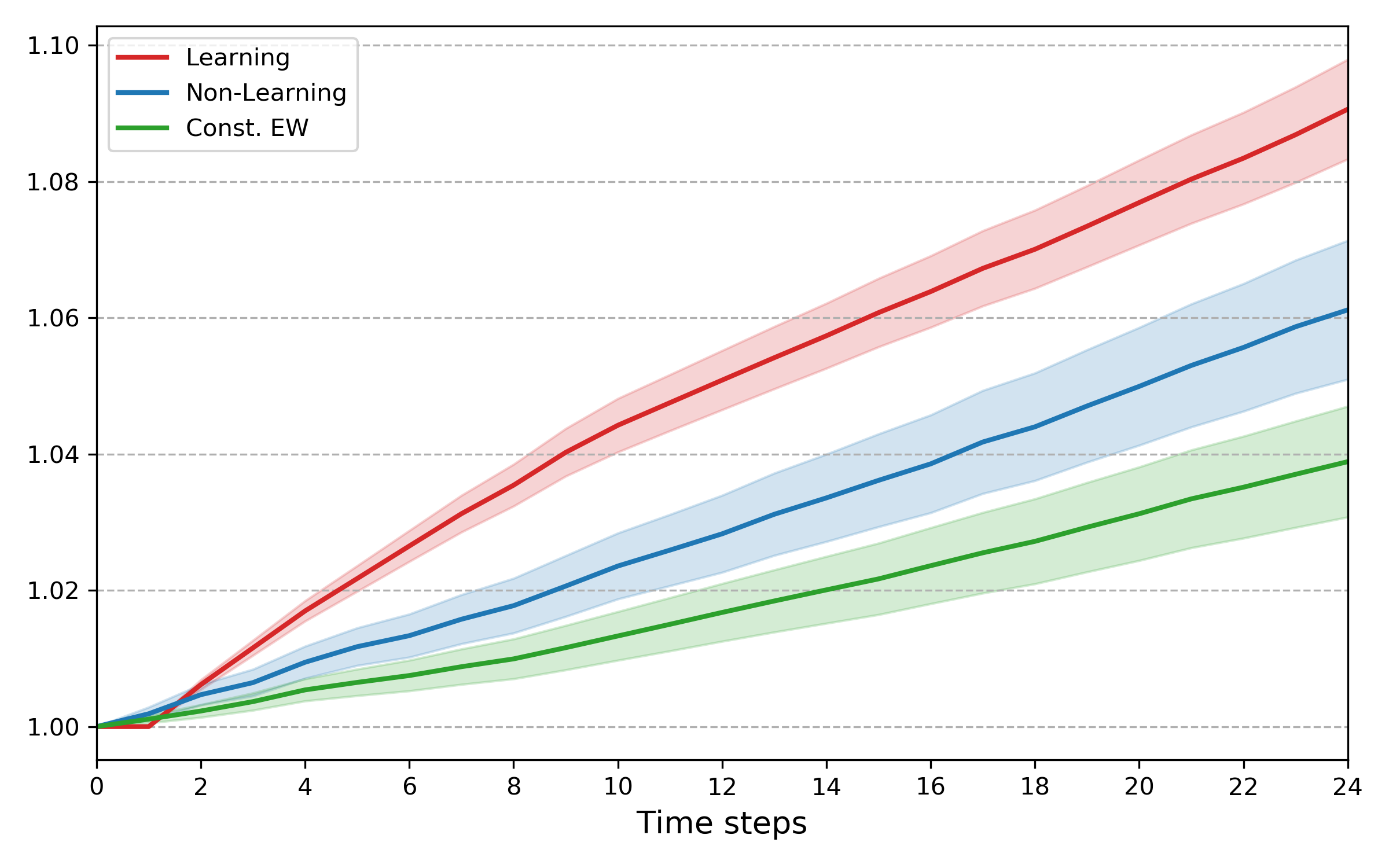}       
\caption{Historical Learning, Non-Learning and constrained EW (Const. EW) levels with a 95\% confidence interval.}\label{Fig:MC_mean_std_Admissible}
\end{figure}

Fig. \ref{Fig:MC_mean_std_Admissible} shows the average historical levels of the three strategies: Learning, Non-Learning and constrained EW. We notice Non-Learning outperforms constrained EW and both have similar confidence intervals. It is not surprising to see that Non-Learning outperforms constrained EW since Non-Learning always bets on Asset 3, the most performing, while constrained EW diversifies the risks equally among the three assets. 

\begin{figure}[H]
	\makebox[1.0 \textwidth][c]{       %centering table
	\resizebox{1.0 \textwidth}{!}{
	\begin{minipage}{0.49\textwidth}
		\centering
		\includegraphics[width=1.05\linewidth]{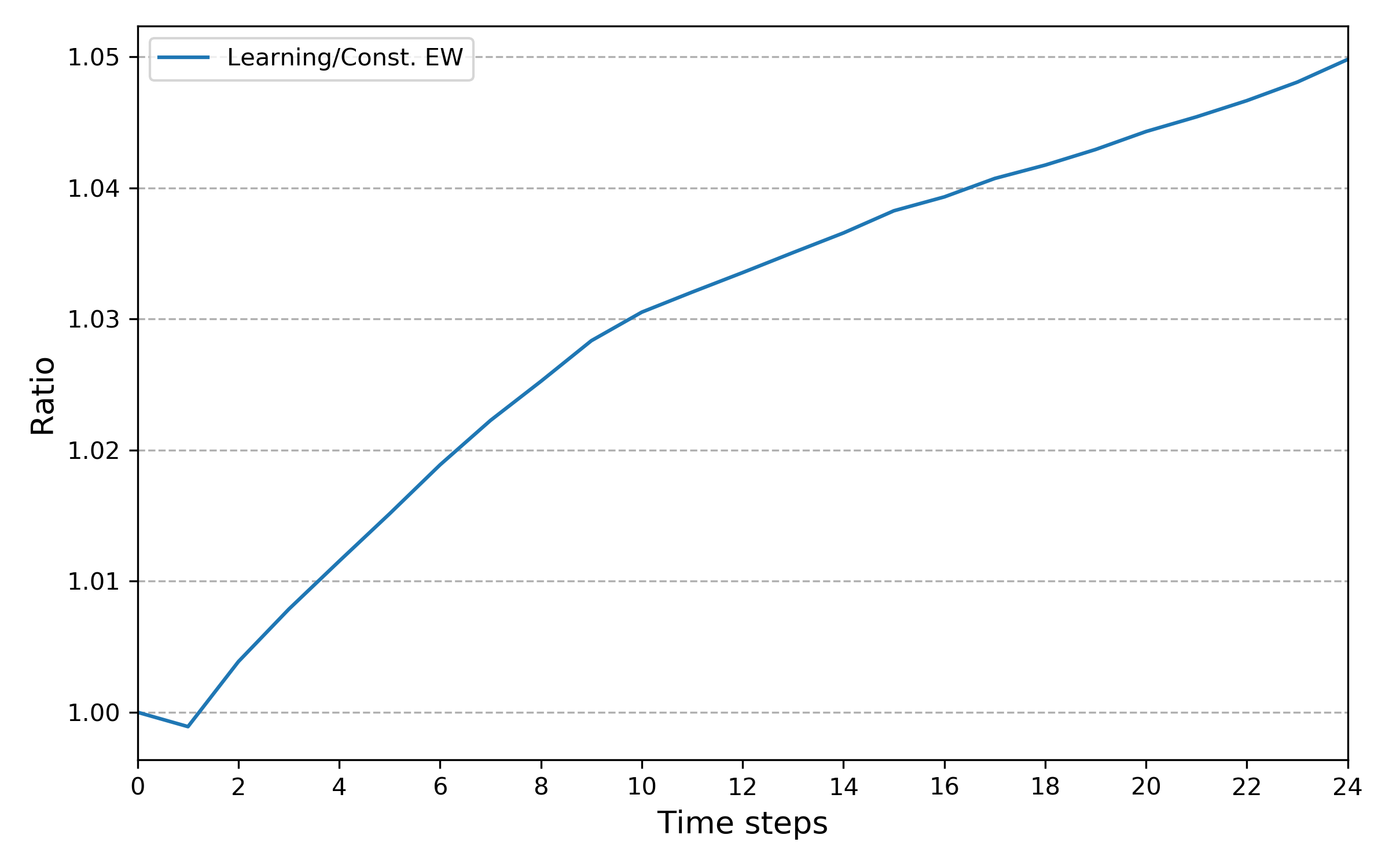}
		\caption{Ratio Learning over constrained EW (Const. EW) according to time.}\label{Fig:MC_Ratio_Learning-EW_const}
	\end{minipage}\hspace{0.1cm}
    \begin{minipage}{0.49\textwidth}
    	\centering
    	\includegraphics[width=1.05\linewidth]{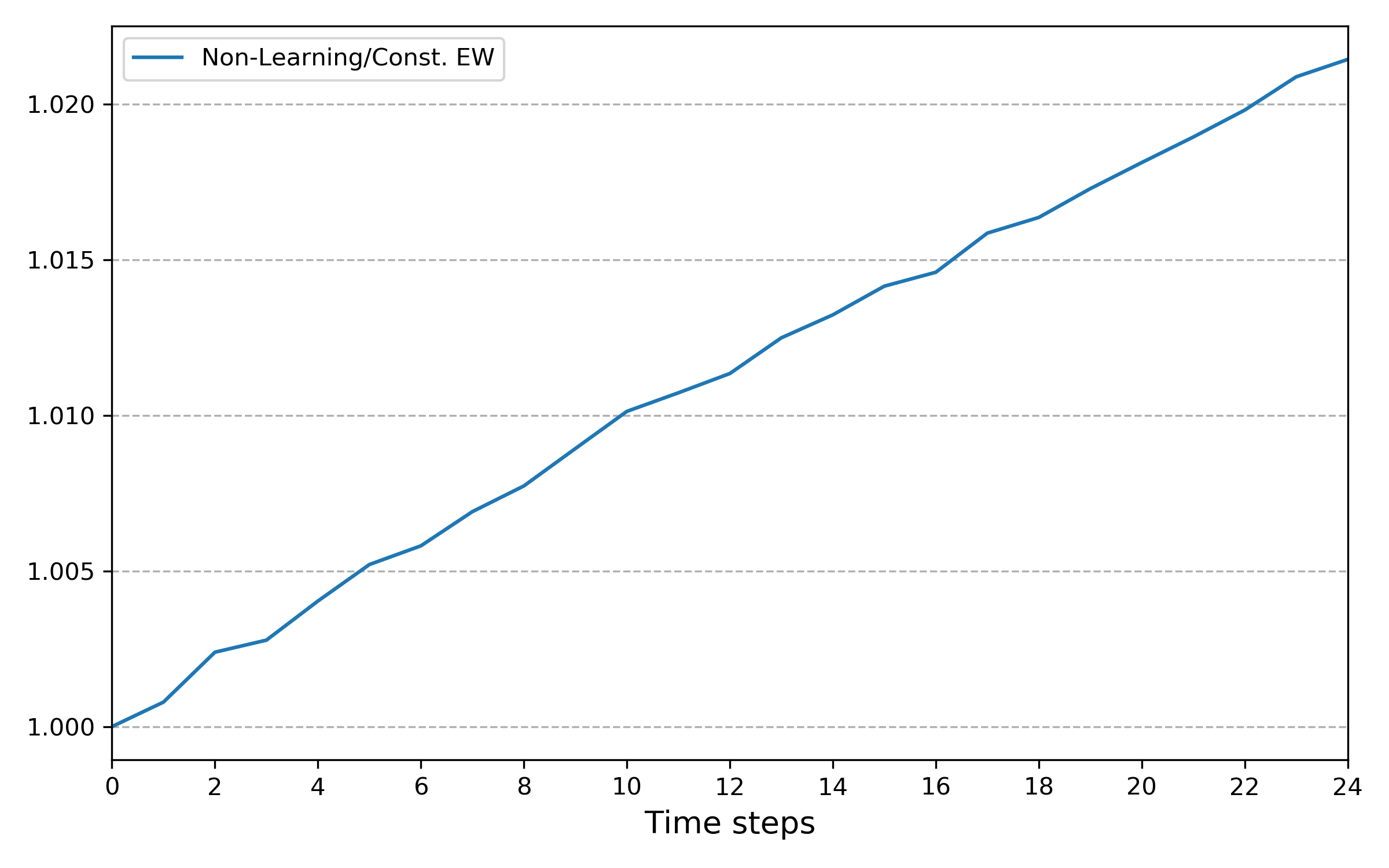}
    	\caption{Ratio Non-Learning over constrained EW (Const. EW) according to time.}\label{Fig:MC_Ratio_Non-learning-EW_const}
    \end{minipage}
     }
     }
\end{figure}

Fig. \ref{Fig:MC_Ratio_Learning-EW_const} shows the ratio of Learning over constrained EW: it depicts the same concave shape as Fig. \ref{Fig:MC_Ratio_Learning-Non-learning}. The outperformance of Non-Learning with respect to constrained EW is plot in Fig. \ref{Fig:MC_Ratio_Non-learning-EW_const} and confirms, on average, the similarity of the two strategies.

Table \ref{tab: Statistics_EW_L_NL} collects relevant statistics for the three strategies. 
Learning clearly surpasses constrained EW: it outperforms by $5.49\%$ while reducing uncertainty on terminal wealth by  $1.92\%$ resulting in an improvement of $182.08\%$ of the Sharpe ratio. Moreover, it better handles maximum drawdown regarding both the average and the worst case, exhibiting an improvement of $3.17\%$ and $10.09\%$ respectively, enhancing the Calmar ratio by $647.56\%$.

The Non-Learning and the constrained EW have similar profiles. Even if Non-Learning outperforms constrained EW by $2.5\%$, it has a higher uncertainty in terminal wealth ($+2.87\%$). This results in similar Sharpe ratios. Maximum drawdown, both on average and considering the worst case are better handled by constrained EW ($-4.70\%$ and $-21.83\%$ respectively) than by Non-Learning ($-6.54\%$ and $-27.18\%$ respectively) thanks to the diversification capacity of constrained EW. The better performance of Non-Learning compensates the better maximum drawdown handling of constrained EW, entailing a better Calmar ratio for Non-Learning $0.98 $ versus $0.82$ for constrained EW.   

\begin{table}[!t]
      \centering 
      \scalebox{0.8}{\begin{tabular}{lccccc} 
        \hline\noalign{\smallskip}
          Statistic & Const. EW & L & NL & L - Const. EW & NL - Const. EW \\
         \noalign{\smallskip}\hline\noalign{\smallskip}       	
         	Avg total performance & 3.85\%   &9.34\%   &6.40\%   &5.49\%   &2.55\%     \\		
        	Std dev. of $X_T$     &13.80\%    &11.88\% &16.67\%  &-1.92\%  &2.87\%     \\      	
        	Sharpe ratio          & 0.28     &0.79     &0.38     &182.08\% &37.63\%   \\       
         	Avg MD     &-4.70\%   &-1.53\%  &-6.54\%  &3.17\%   &-1.84\%    \\  
         	Worst MD    &-21.83\%  &-11.74\%  &-27.18\% &10.09\%  &-5.34\%    \\     	
         	Calmar ratio          &0.82      &6.12     & 0.98    &647.56\% &-19.56\%   \\			
        \noalign{\smallskip}\hline\noalign{\smallskip}
      \end{tabular}}    
      \caption{Performance metrics: Constrained EW (Const. EW) vs Learning (L) and Non-Learning (NL). The difference for ratios are computed as relative improvement.} 
      \label{tab: Statistics_EW_L_NL}
\end{table}

\subsubsection{Non-learning and Merton strategies} \label{subsec: NL_and_Merton_strategies}
\begin{figure}[H]
     \centering
     \includegraphics[scale=0.4]{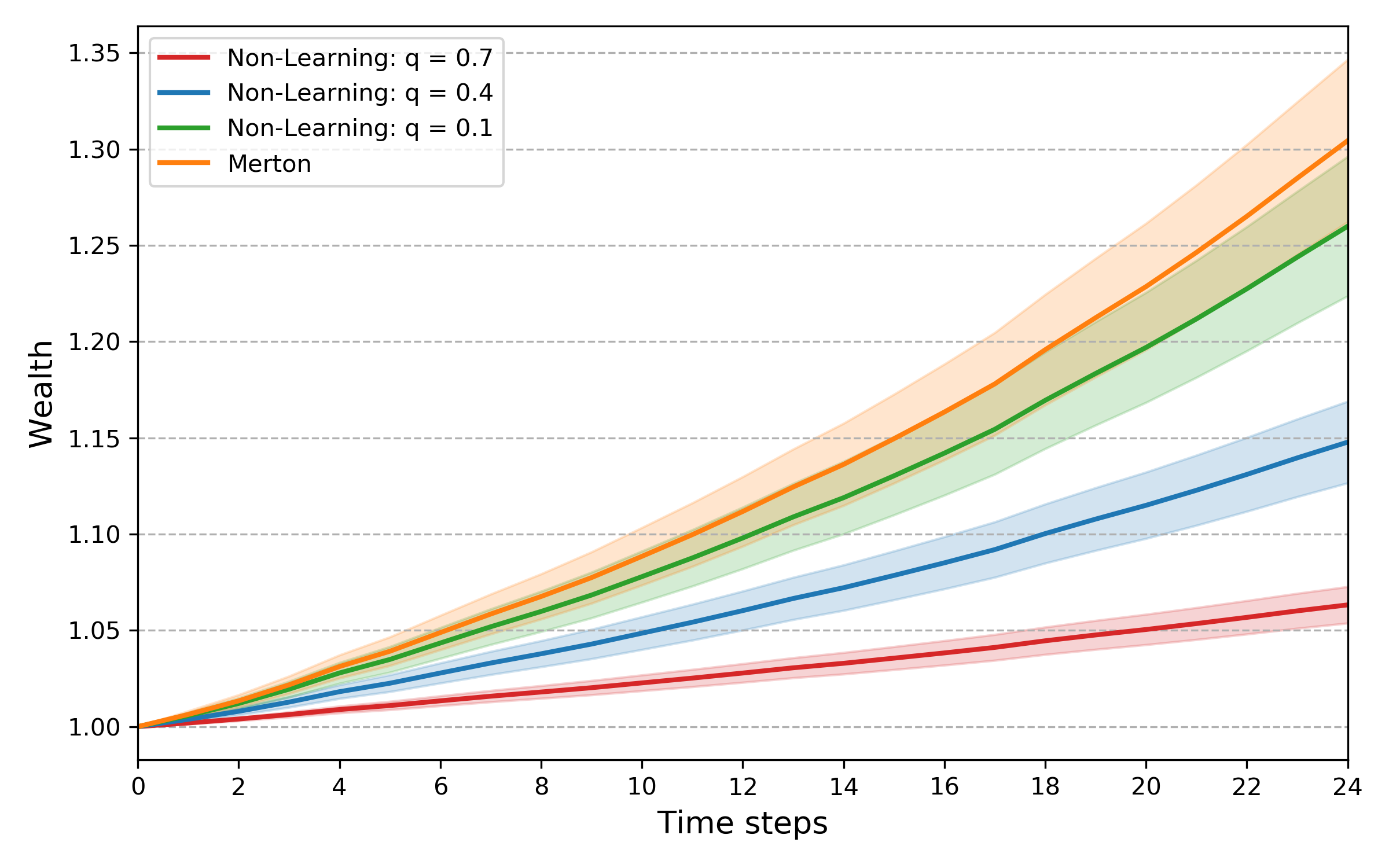}
     \caption{Wealth curves resulting from the Merton strategy and the non-learning strategy for different values of $q$.}\label{Fig:MC_mean_std_NL_Merton_conv}
\end{figure}

We numerically analyze the impact of the drawdown parameter $q$, and compare the non-learning strategies (assuming that the drift is equal to $b_0$), with the constrained Merton strategy as described in Remark \ref{remMerton}. Fig. \ref{Fig:MC_mean_std_NL_Merton_conv} confirms  that when the loss aversion parameter $q$ goes to zero, 
the non-learning strategy approaches the Merton strategy. 

\begin{figure}[H]
	\makebox[1.0 \textwidth][c]{       %centering table
	\resizebox{1.0 \textwidth}{!}{
     \includegraphics[scale=.40]{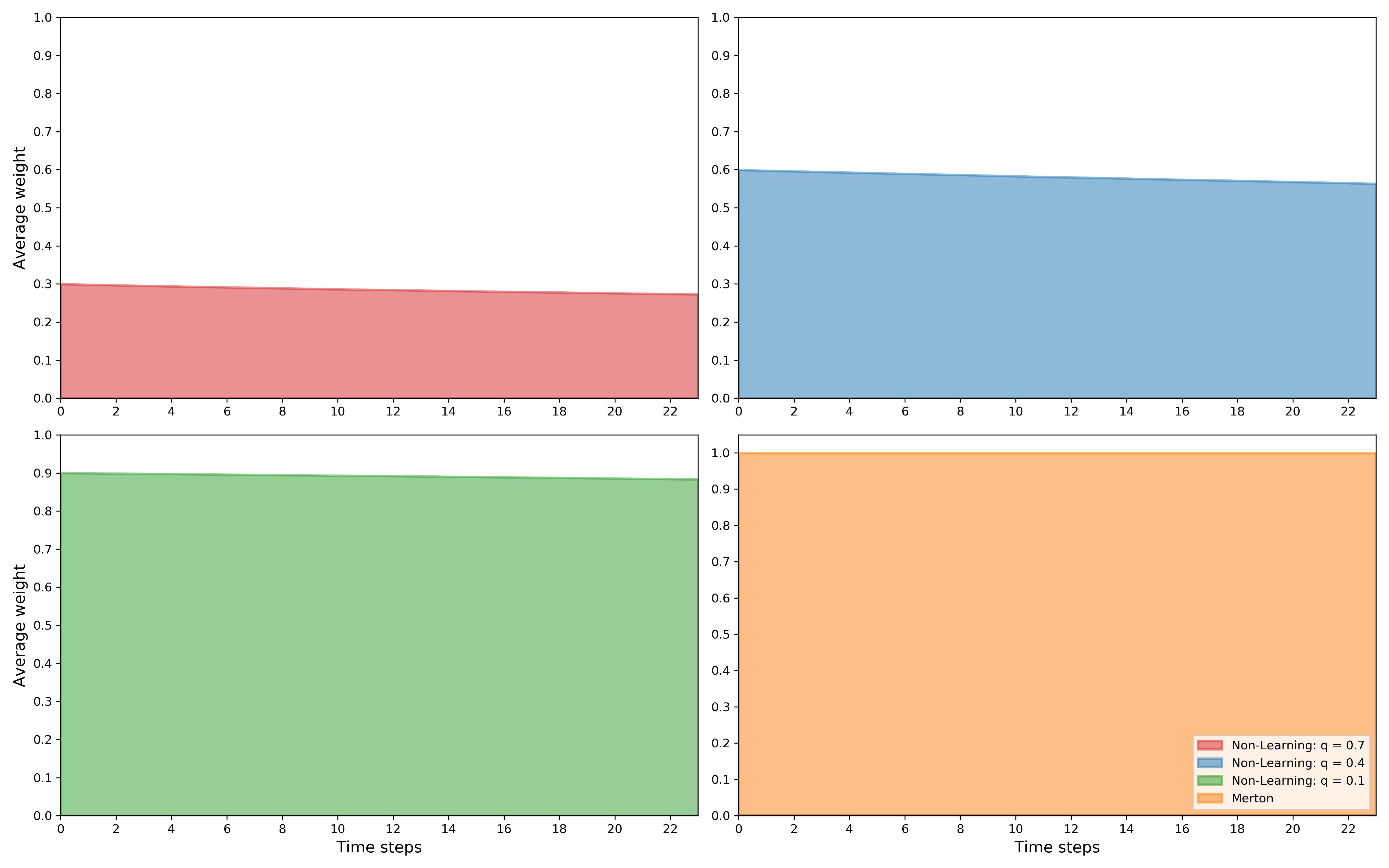}
     }
     }
     \caption{Asset $3$ average weights of the non-learning strategies with $q \in \{0.7,0.4,0.1\}$ and the Merton strategy.}\label{Fig:Weights_Area_Asset_3}
\end{figure}

In terms of assets' allocation, the Merton strategy saturates the constraint only by investing in the asset with the highest expected return, Asset $3$, while the non-learning strategy adopts a similar approach and invests at full capacity in the same asset. To illustrate this point, we easily see that the areas at the top and bottom-left corner converge to the area at the bottom-right corner of Fig. \ref{Fig:Weights_Area_Asset_3}.

As $q$ vanishes, we observe evidence of the convergence of the Merton and the non-learning strategies, materialized by a converging allocation pattern and resulting wealth trajectories. It should not be surprising since both have in common not to  learn from incoming information conveyed by the prices. 

\subsection{Sensitivities analysis} \label{subsec: Sens_anal}
In this subsection, we study the effect of changes in the uncertainty about the beliefs of $B$. These beliefs take the form of an estimate $b_0$ of $B$,  and a degree of uncertainty about this estimate, the covariance of  $\covB$ of $B$. For the sake of simplicity, we design $\covB$ as a diagonal matrix whose diagonal entries are variances representing the confidence the investor has in her beliefs about the drift. To easily model a change in $\covB$, we define the modified covariance matrix $\tilde{\Sigma}$ as
\beqs
	\tilde{\Sigma}_{unc}:= unc * \covB,
\enqs
where $unc >0$. From now on, the prior of $B$ is $\mathcal{N}(b_0,\tilde{\Sigma}_{unc})$.

A higher value of $unc$ means a higher uncertainty materialized by a lower confidence in the prior estimate of the expected return of $B$, $b_0$. We consider learning strategies with values of $unc \in \{1/6,\;1,\;3,\;6,\;12 \}$. The value $unc=1$ was used for Learning in Subsection \ref{subsec: num_res}. \\

\begin{figure}[H]
     \centering
     \includegraphics[scale=0.4]{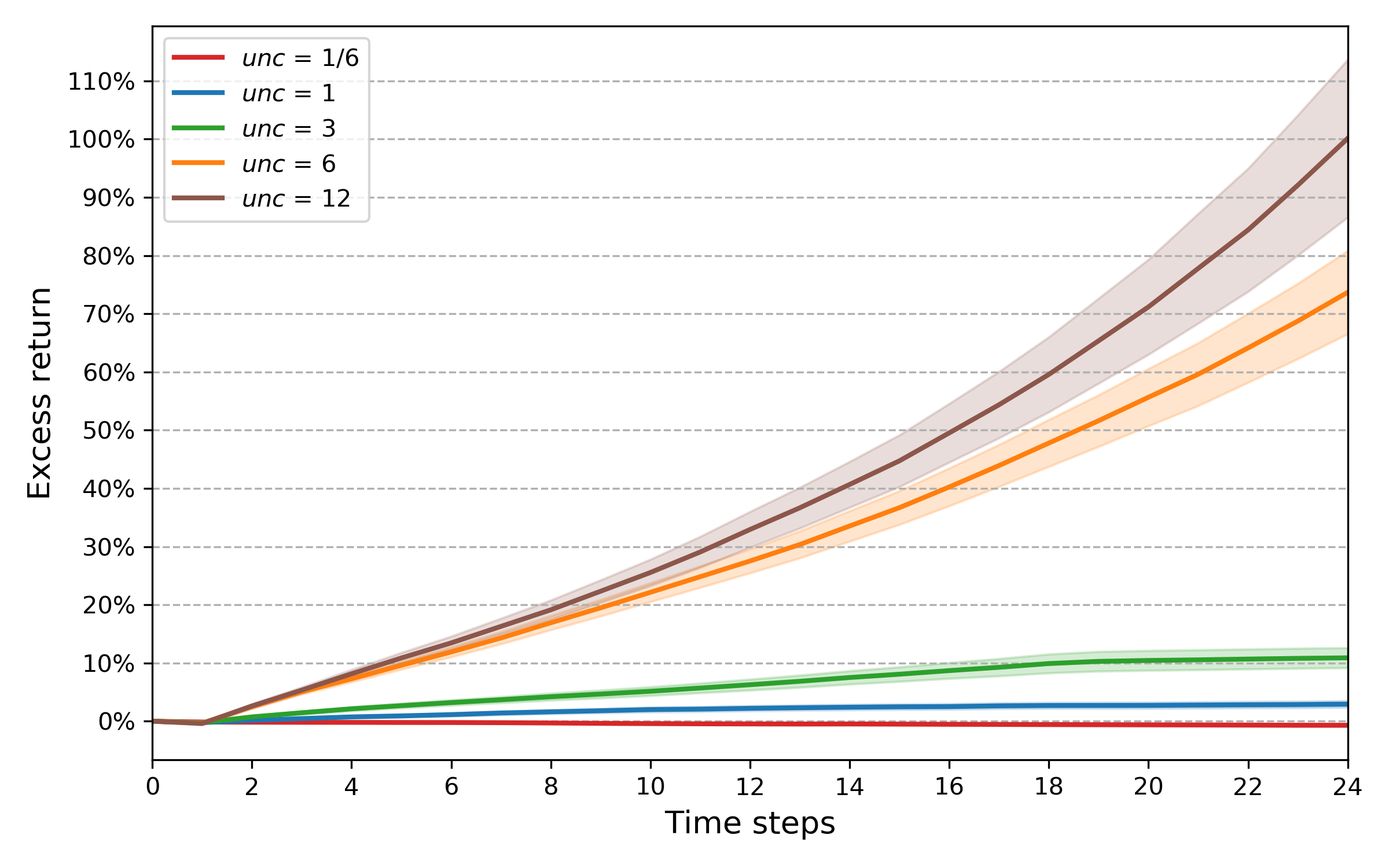}
     \caption{Excess return of Learning over Non-Learning with a 95\% confidence interval for different levels of uncertainty.} \label{Fig:XS_ret.png}
\end{figure}

Equation \eqref{eq: ret_process} implies that the returns' probability distribution depends upon $unc$. It implies that for each value of $unc$, we need to compute both Learning and Non-Learning  on the returns sample drawn from the same probability law to make relevant comparisons.\\
Therefore, from a sample of a thousand returns paths' draws, we plot in Fig. \ref{Fig:XS_ret.png} the average curves of the excess return of Learning over its associated Non-Learning, for different values of the uncertainty parameter \emph{unc}. 

Looking at Fig. \ref{Fig:XS_ret.png}, we notice that when uncertainty about $b_0$ is low, i.e. $unc=1/6$, Learning is close to Non-Learning and unsurprisingly the associated excess return is small. Then, as we increase the value of \emph{unc} the curves steepen increasingly showing the effect of learning in generating excess return.\\

\begin{figure}[H]
	\begin{minipage}{0.49\textwidth}
	\centering
	\includegraphics[scale=0.5]{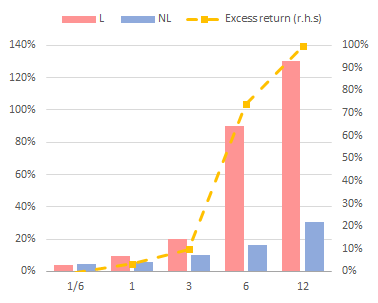}
	\caption{Average total performance of Learning (L) and Non-Learning (NL), and excess return, for \emph{unc} $\in \{1/6,1,3,6,12\}$.}\label{Fig:XS_price_unnorm_excel.png}
	\end{minipage}\hfill
   \begin{minipage}{0.49\textwidth}
     \centering
     \includegraphics[scale=0.5]{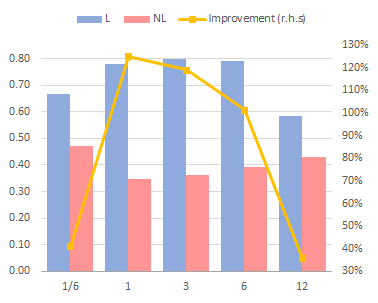}
     \caption{Sharpe ratio of terminal wealth of Learning (L) and Non-Learning (NL), and relative improvement, for \emph{unc} $\in \{1/6,1,3,6,12\}$.}\label{Fig:Risk_reward_excel.png}
     \end{minipage}
\end{figure}

Table \ref{tab: Statistics_L_NL_unc} summarises key statistics for the ten strategies computed in this section. When $\emph{unc}=1/6$, Learning underperforms Non-Learning. This is explained by the fact that Non-Learning has no doubt about $b_0$ and knows Asset 3 is the best performing asset acoording to its prior, whereas Learning, even with low uncertainty, needs to learn it generating a lag which explains the underperformance on average. For values of $\emph{unc}\geq 1$ Learning outperforms Non-learning increasingly, as can be seen on Fig. \ref{Fig:XS_price_unnorm_excel.png}, at the cost of a growing standard deviation of terminal wealth. \\

The Sharpe ratio of terminal wealth is higher for Learning than for Non-Learning for any value of $\emph{unc}$. Nevertheless, an interesting fact is that the ratio rises from $unc=1/6$ to $unc=1$, then reaches a level close to $0.8$ for values of $\emph{unc} = 1,3,6$ then decreases when $\emph{unc} = 12$. 

\begin{table}[!t]
\makebox[1 \textwidth][c]{       %centering table
\resizebox{1. \textwidth}{!}{
\begin{tabular}{lcccccccccc} 
      	\cline{2-11}
      	\multicolumn{1}{c}{}& \multicolumn{2}{c}{${\emph{unc} = 1/6}$} & \multicolumn{2}{c}{${\emph{unc} = 1}$} & \multicolumn{2}{c}{${\emph{unc} = 3}$}& \multicolumn{2}{c}{${\emph{unc} = 6}$} & \multicolumn{2}{c}{${\emph{unc} = 12}$}\\
       \hline\noalign{\smallskip}
        Statistic & L & NL & L & NL & L & NL & L & NL & L & NL\\
         \noalign{\smallskip}\hline\noalign{\smallskip}       	
         	Avg total performance   & 3.87\%  & 4.35\%   & 9.45\%  & 6.00\%   & 19.96\% & 10.25\%  & 90.03\%  & 16.22\%  & 130.07\%  & 30.44\%  \\
        	Std dev. of $X_T$           & 5.81\%  & 9.22\%   & 12.10\% & 17.28\%  & 25.01\% & 28.18 \% & 113.69\% & 41.24\%  & 222.77\%  & 70.84\%  \\      	
        	Sharpe ratio       & 0.67    & 0.47     & 0.78    & 0.35     & 0.80    & 0.36     & 0.79     & 0.39     & 0.58      & 0.43    \\       
         	Avg MD  & -2.51\% & -5.21\%    &-1.40\%  & -6.78\%  & -1.90\% & -8.40\%  & -2.68\%  & -10.14\% & -3.58\%   &  -11.35\%\\   
         	Worst MD & -7.64\% & -17.88\% &-5.46\%  & -24.01\% & -7.99\% & -26.68\% & -15.62\% & -29.22\% & -16.98\%  &  -29.47\%\\       	
         	Calmar ratio       & 1.54    & 0.83     & 6.77    & 0.89     & 10.49   & 1.22     & 33.65    & 1.60     & 36.32     &  2.68   \\			
        \noalign{\smallskip}\hline\noalign{\smallskip}
      \end{tabular}    
      }
      }
      \caption{Performance and risk metrics: Learning (L) vs Non-Learning (NL) for different values of uncertainty $\emph{unc}$.} 
      \label{tab: Statistics_L_NL_unc}
\end{table}

\noindent This phenomenon is more visible on Fig. \ref{Fig:Risk_reward_excel.png} that displays the Sharpe ratio of terminal wealth of Learning and Non-Learning according to the values of $\emph{unc}$, and the associated relative improvement. Clearly, looking at Figures \ref{Fig:XS_price_unnorm_excel.png} and \ref{Fig:Risk_reward_excel.png}, we remark that while increasing $unc$ gives more excess return, too high values of $unc$ in the model turn out to be a drag as far as Sharpe ratio improvement is concerned.

For any value of $\emph{unc}$, Learning handles maximum drawdown significantly better than Non-Learning whatever  it is the average or the worst. This results in a better performance per unit of average maximum drawdown (Calmar ratio), for Learning. We also see that the maximum drawdown constraint is satisfied for every strategies of the sample and for any value of $unc$ since the worst maximum drawdown is always above $-30\%$, the lowest admissible value with a loss aversion parameter $q$ set at $0.7$.

\begin{figure}[H]
	\centering
     \includegraphics[scale=0.6]{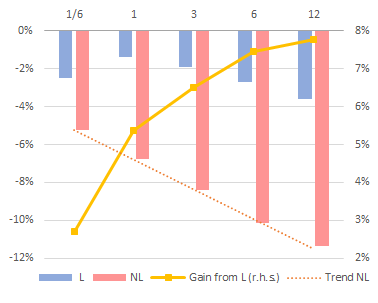}
     \caption{Average maximum drawdown of Learning (L) and Non-Learning (NL) and the gain from learning for \emph{unc} $\in \{1/6,1,3,6,12\}$.}\label{Fig:XS_MD_excel.png}
\end{figure}

\noindent Fig. \ref{Fig:XS_MD_excel.png} reveals how the average maximum drawdown behaves regarding the level of uncertainty. Non-Learning maximum drawdown behaves linearly with uncertainty: the wider the range of possible values of $B$ the higher the maximum drawdown is on average. It emphasizes its inability  to adapt to an environment in which the returns have different behaviors compared to their expectations. Learning instead, manages to keep a low maximum drawdown for any value of $\emph{unc}$. Given the previous remarks, it is obvious that the gain in maximum drawdown from learning grows with the level of uncertainty.\\

\begin{figure}[H]
     \makebox[1.0 \textwidth][c]{       %centering table
	  \resizebox{1.0 \textwidth}{!}{
     \includegraphics[width=1.05\linewidth]{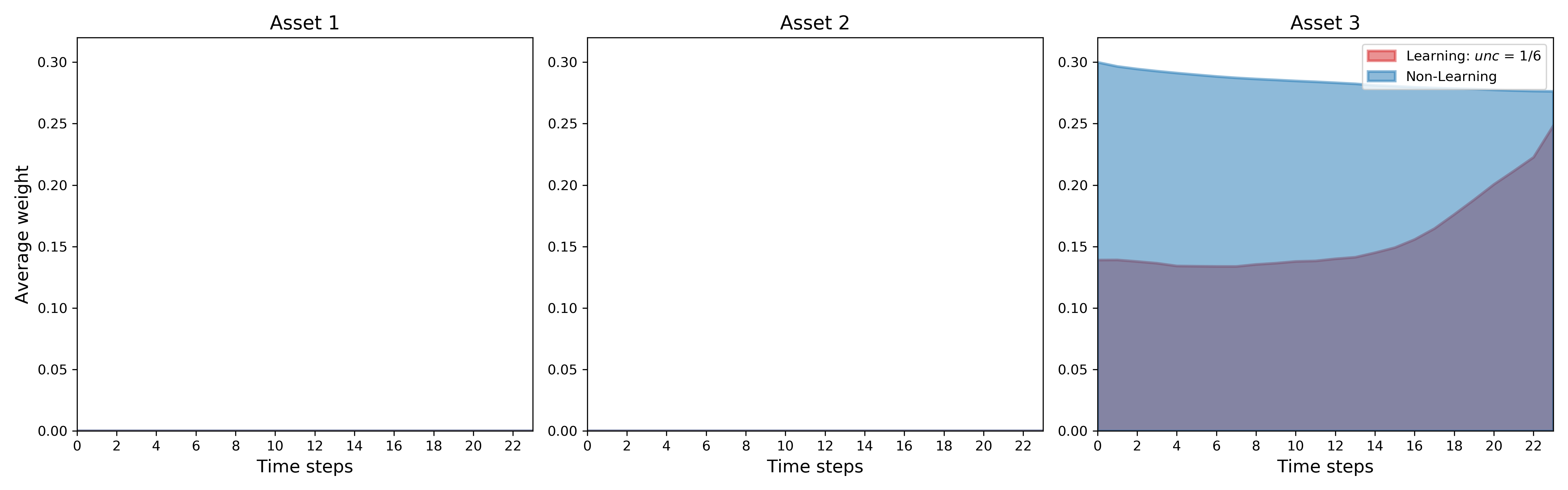}
     }
     }
     \caption{Learning and Non-Learning historical assets' allocations with $unc=1/6$.}\label{Fig:Weights_Area_L_NL_16}
\end{figure}

\begin{figure}[H]
	\makebox[1.0 \textwidth][c]{       %centering table
	\resizebox{1.0 \textwidth}{!}{
     \includegraphics[width=1.05\linewidth]{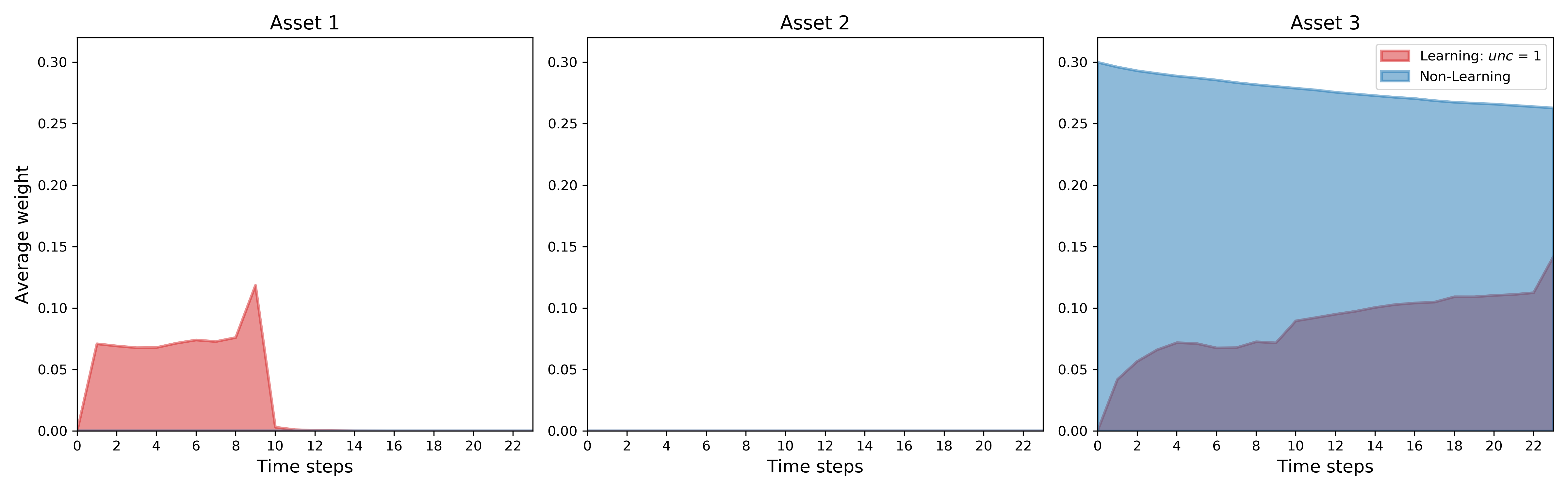}
     }
     }
     \caption{Learning and Non-Learning historical assets' allocations with $unc=1$.}\label{Fig:Weights_Area_L_NL_1}
\end{figure}

\begin{figure}[H]
     \makebox[1.0 \textwidth][c]{       %centering table
	\resizebox{1.0 \textwidth}{!}{
     \includegraphics[width=1.05\linewidth]{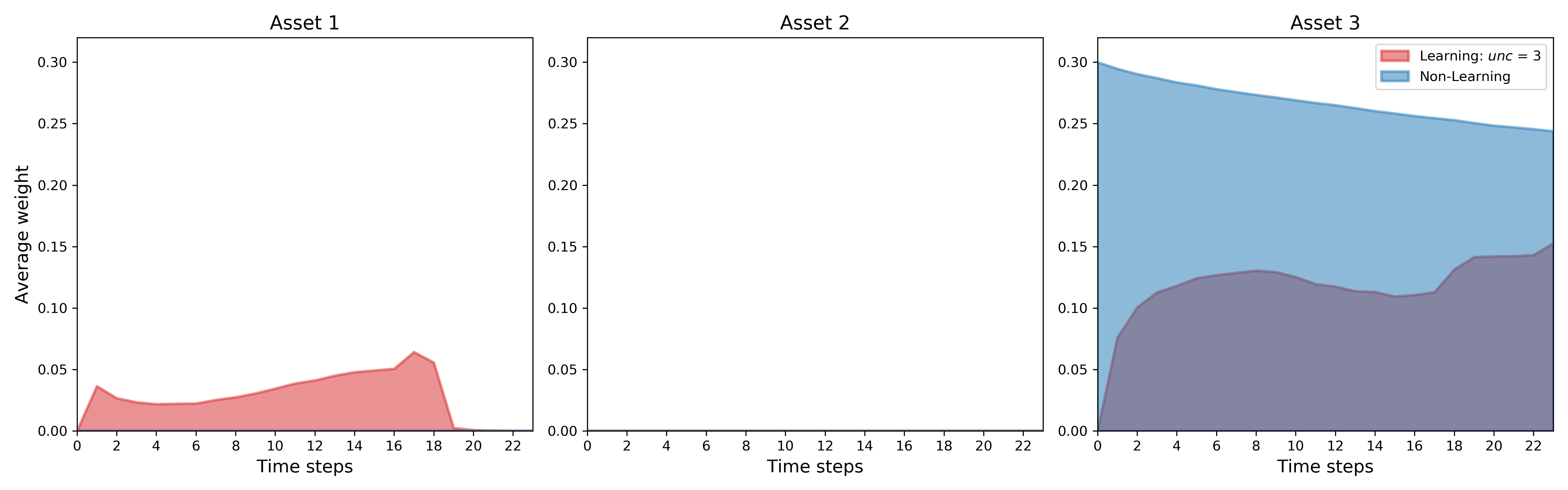}
      }
      }
     \caption{Learning and Non-Learning historical assets' allocations with $unc=3$.}\label{Fig:Weights_Area_L_NL_3}
\end{figure}

\begin{figure}[H]
     \makebox[1.0 \textwidth][c]{       %centering table
	\resizebox{1.0 \textwidth}{!}{
     \includegraphics[width=1.05\linewidth]{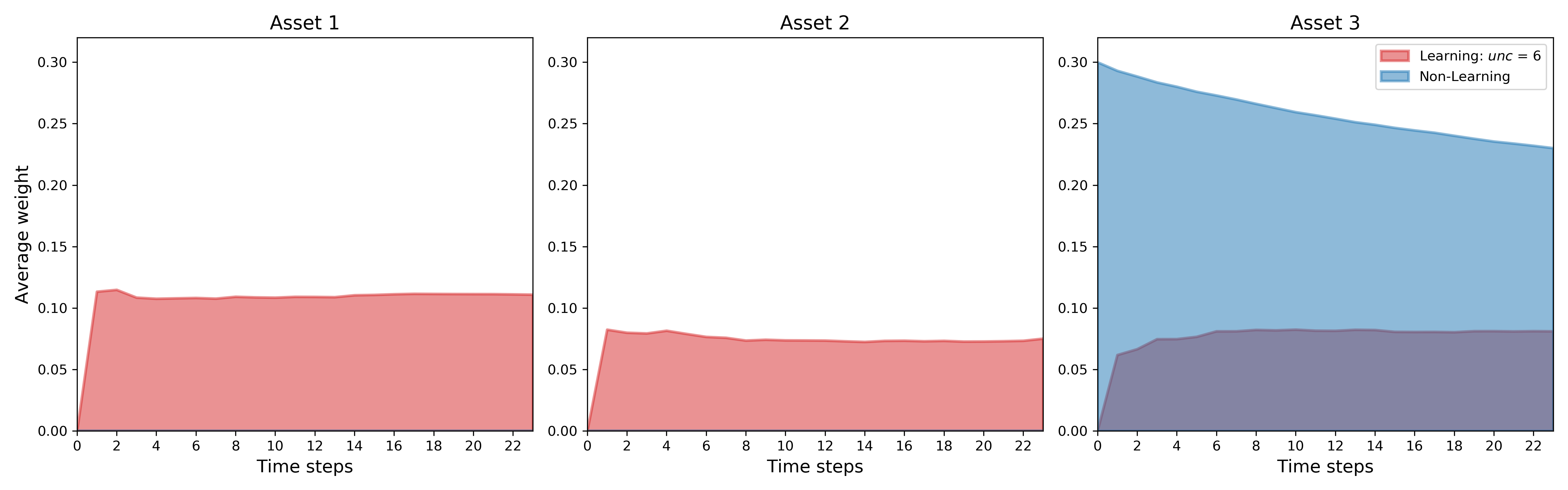}
     }
     }
     \caption{Learning and Non-Learning historical assets' allocations with $unc=6$.}\label{Fig:Weights_Area_L_NL_6}
\end{figure}

\begin{figure}[H]
     \makebox[1.0 \textwidth][c]{       %centering table
	 \resizebox{1.0 \textwidth}{!}{
     \includegraphics[width=1.05\linewidth]{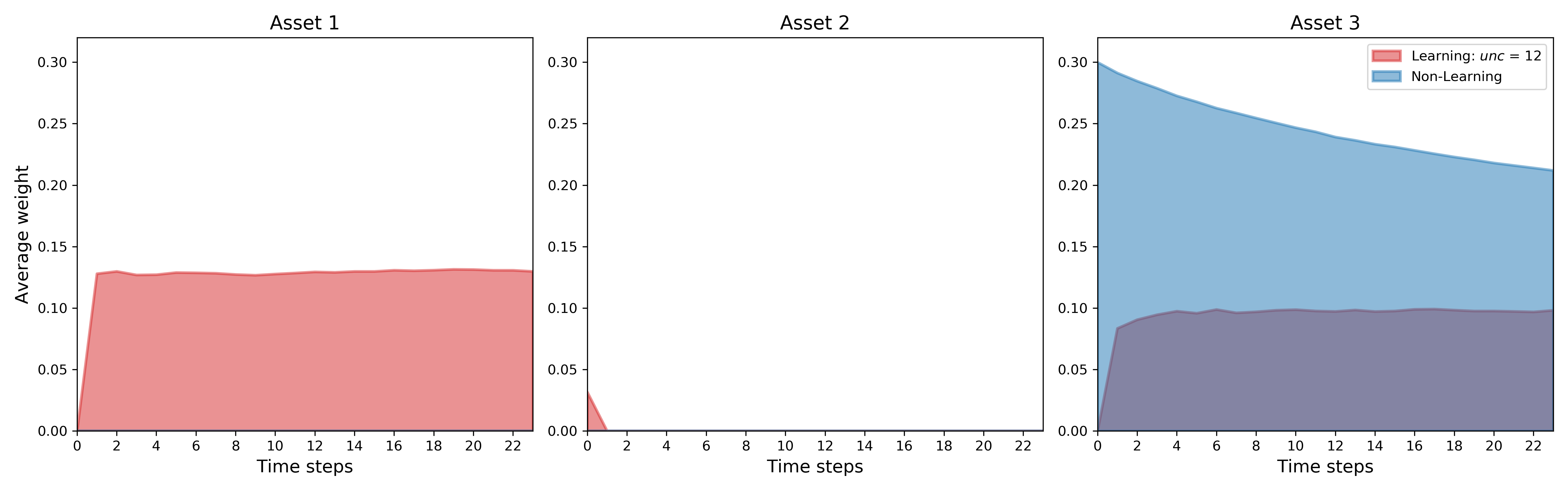}
     }
     }
     \caption{Learning and Non-Learning historical assets' allocations with $unc=12$.}\label{Fig:Weights_Area_L_NL_12}
\end{figure}

Figures \ref{Fig:Weights_Area_L_NL_16}-\ref{Fig:Weights_Area_L_NL_12} represent portfolio allocations averaged over the simulations. They depict, for each value of the uncertainty parameter $unc$, the average proportion of wealth invested, in each of the three assets, by Learning and Non-Learning. The purpose is not to compare the graphs with different values of $unc$ since the allocation is not performed on the same sample of returns. Rather, we can identify trends that are typically differentiating Learning from Non-Learning allocations.\\

Since the maximum drawdown constraint is satisfied by the capped sum of total weights that can be invested, the allocations of both Learning and Non-Learning are mainly based on the expected returns of the assets.\\
Non-Learning, by definition, does not depend on the value of the uncertainty parameter. Hence, no matter the value of $unc$, its allocation is easy to characterize since it saturates its constraint investing in the asset that has the best expected return according to the prior. In our setup, Asset $3$ has the highest expected return, so Non-Learning invests only in it and saturates its constraint of roughly $30\%$ during all the investment period. The slight change of the average weight in Asset $3$ comes from $\rho$, the ratio wealth over maximum wealth, changing over time.\\

Unlike Non-Learning, depending of the value of $unc$, Learning can perform more sophisticated allocations because it can adjust the weights according to the incoming information. Nonetheless, in Fig. \ref{Fig:Weights_Area_L_NL_16}, when $unc$ is low, Learning and Non-Learning look similar regarding their weights allocation since both strategies invest, as of time $0$, a significant proportion of their wealth only in Asset $3$.
\\
On the right panel of Fig. \ref{Fig:Weights_Area_L_NL_16}, the progressive increase in the weight of Asset $3$ illustrates the learning process. As time goes by, Learning progressively increases the weight in Asset $3$ since it has the highest expected return. It also explains why Learning underperforms Non-Learning for low values of $unc$; contrary to Non-Learning which invests at full capacity in Asset $3$, Learning needs to learn that Asset $3$ is the optimal choice.

However, as uncertainty increases, Learning and Non-Learning strategies start differentiating. When $unc\geq 1$, Learning invests little, if any, at time $0$. In addition, an increase in $unc$ allows the inital drift to lie in a wider range and generates investment opportunities for Learning. This explains why Learning invests in Asset $1$ when $unc={1,\;3,\;6,\;12}$ although the estimate $b_0$ for this asset is lower than for Asset $3$. In Fig. \ref{Fig:Weights_Area_L_NL_6}, we see that Learning even invests in Asset $2$ which has the lowest expected drift.

\begin{figure}[H]
     \centering
     \includegraphics[width=1.05\linewidth]{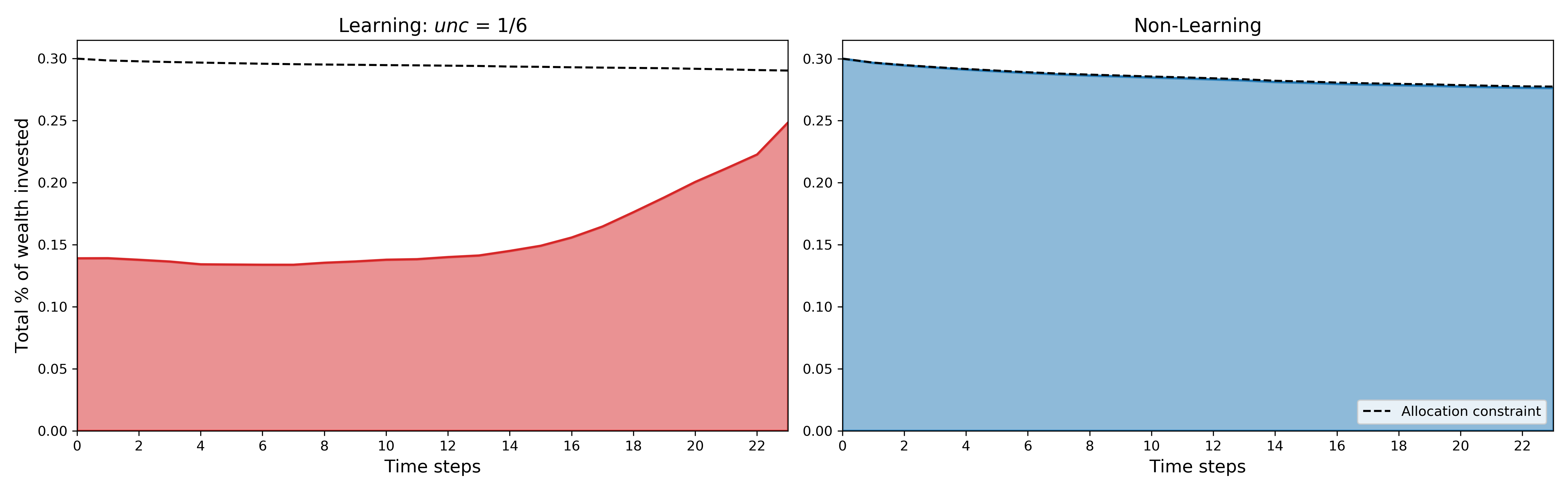}
     \caption{Historical total allocations of Learning and Non-Learning with $unc=1/6$.}\label{Fig:Weights_Sums_L_NL_16}
\end{figure}

\begin{figure}[H]
     \centering
     \includegraphics[width=1.05\linewidth]{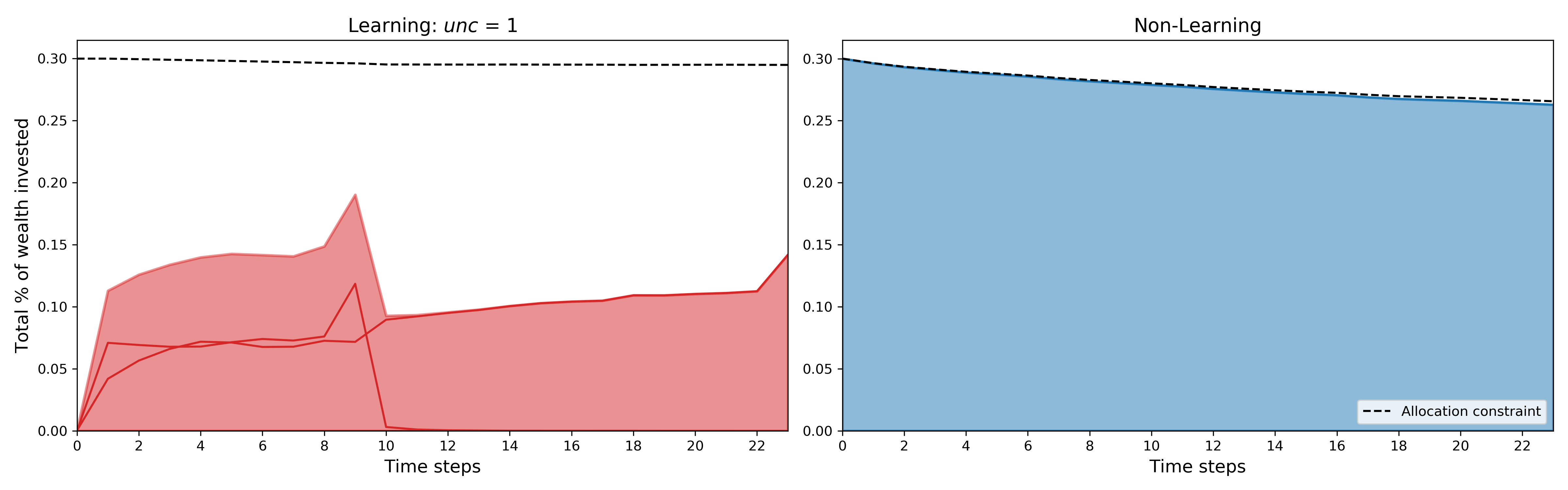}
     \caption{Historical total allocations of Learning and Non-Learning with $unc=1$.}\label{Fig:Weights_Sums_L_NL_1}
\end{figure}

\begin{figure}[H]
     \centering
     \includegraphics[width=1.05\linewidth]{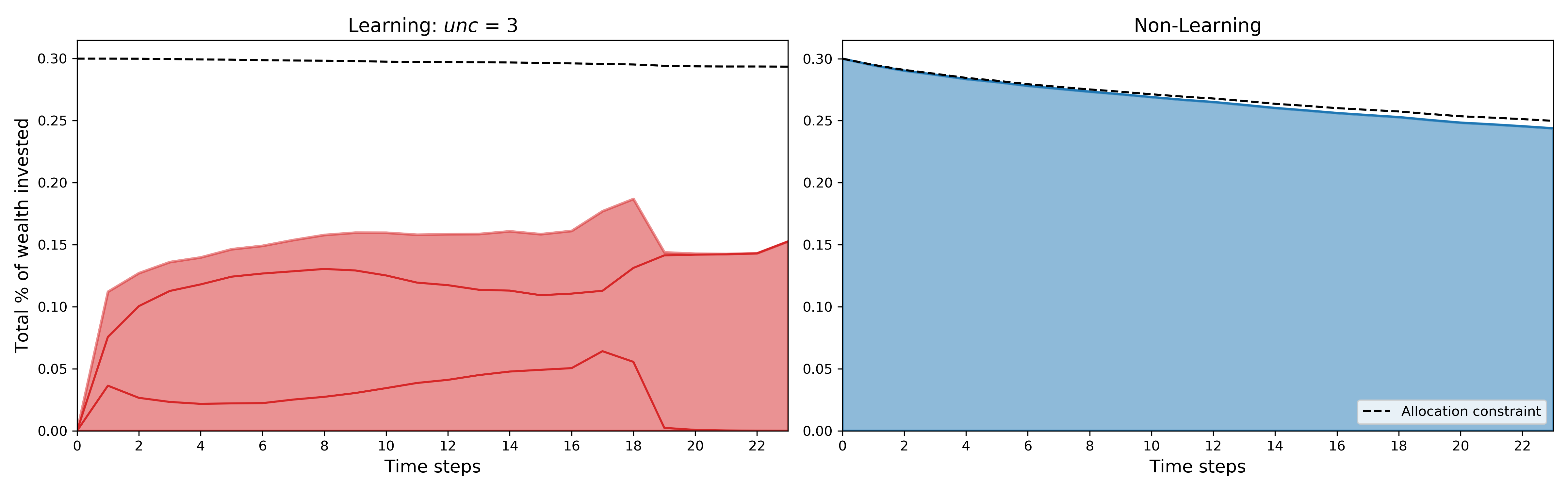}
     \caption{Historical total allocations of Learning and Non-Learning with $unc=3$.}\label{Fig:Weights_Sums_L_NL_3}
\end{figure}

\begin{figure}[H]
     \centering
     \includegraphics[width=1.05\linewidth]{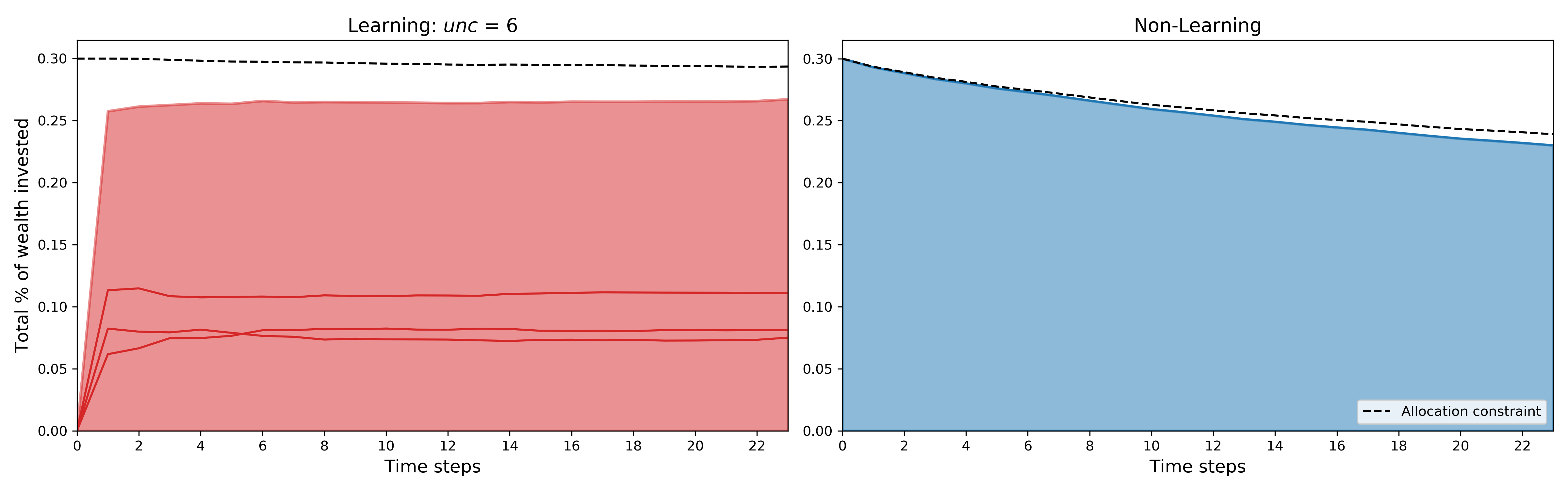}
     \caption{Historical total allocations of Learning and Non-Learning with $unc=6$.}\label{Fig:Weights_Sums_L_NL_6}
\end{figure}

\begin{figure}[H]
     \centering
     \includegraphics[width=1.05\linewidth]{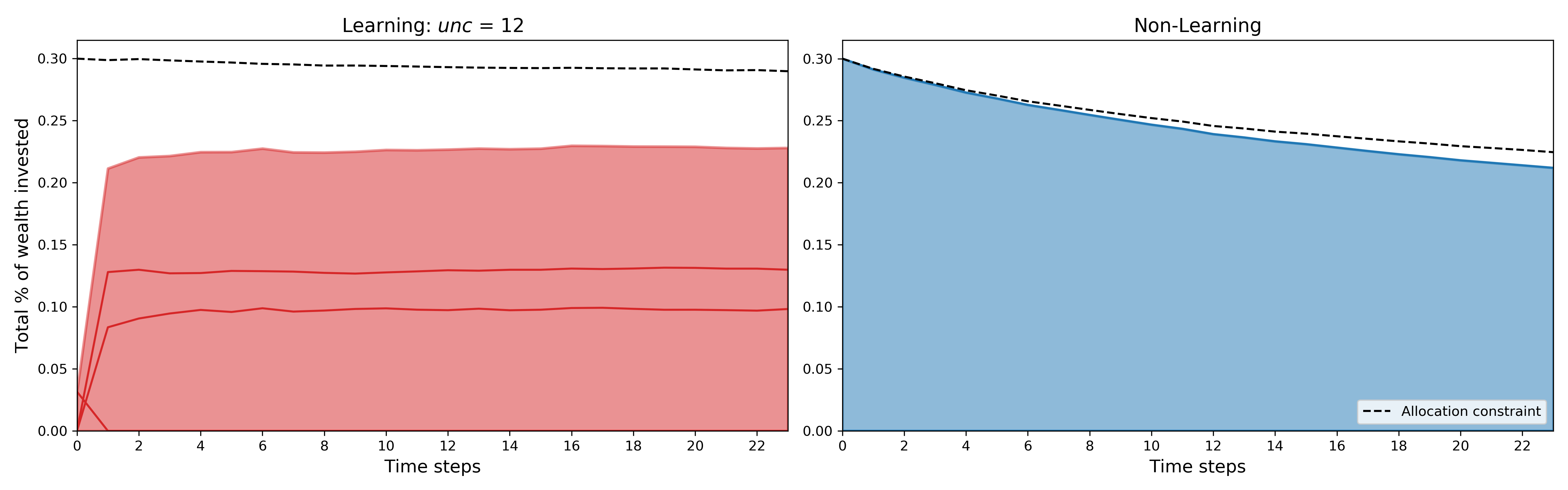}
     \caption{Historical total allocations of Learning and Non-Learning with $unc=12$.}\label{Fig:Weights_Sums_L_NL_12}
\end{figure}

Figures \ref{Fig:Weights_Sums_L_NL_16}-\ref{Fig:Weights_Sums_L_NL_12} illustrate the historical total percentage of wealth allocated for Learning and Non-Learning with different levels of uncertainty. As seen previously, Non-Learning has fully invested in Asset $3$ for any value of $unc$.  

Moreover, Learning has always less investment that Non-Learning for any level of uncertainty. It suggests that Learning yields a more cautious strategy than Non-Learning. This fact, in addition to its wait-and-see approach at time $0$ and its ability to better handle maximum drawdown, makes Learning a safer and more conservative strategy than Non-Learning. This can be seen in Fig. \ref{Fig:Weights_Sums_L_NL_16}, where both Learning and Non-Learning have invested in Asset $3$, but not at the same pace. Non-Learning goes fully in Asset $3$ at time $0$, whereas Learning increments slowly its weight in Asset $3$ reaching $25\%$ at the final step. When $unc$ is low, there is no value added to choose Learning over Non-Learning from a performance perspective. Nevertheless, Learning allows for a better management of risk as Table \ref{tab: Statistics_L_NL_unc} exhibits.\\

As $unc$ increases, in addition to being cautious, Learning mixes allocation in different assets, see Figures \ref{Fig:Weights_Sums_L_NL_1}-\ref{Fig:Weights_Sums_L_NL_12}, while Non-Learning is stuck with the highest expected return asset. 

Learning is able to be opportunistic and changes its allocation given the prices observed. For example in Fig. \ref{Fig:Weights_Sums_L_NL_1}, Learning starts investing in Asset $1$ and $3$ at time $1$ and stops at time $12$ to weigh Asset $1$ while keeping Asset $3$. Similar remarks can be made for Fig. \ref{Fig:Weights_Sums_L_NL_3}, where Learning puts non negligeable weights in all three risky assets for $unc=6$ in Fig. \ref{Fig:Weights_Sums_L_NL_6}.

\section{Conclusion}

We have studied a  discrete-time portfolio selection problem by taking into account both  drift uncertainty and maximum drawdown constraint. The dynamic programming equation has been derived in the general case thanks to a specific change of measure. More explicit results have been provided in the Gaussian case using the Kalman filter. Moreover, a change of variable has reduced the dimensionality of the problem in the case of CRRA utility functions. Next, we have provided extensive numerical results in the Gaussian case with CRRA utility functions using recent deep neural network techniques. Our numerical analysis has clearly shown and quantified the better risk-return profile of the learning strategy versus the non-learning one. Indeed, besides outperforming the non-learning strategy, the learning one provides a significantly lower standard deviation of terminal wealth and a better controlled maximum drawdown. Confirming the results established in \cite*{DeFranco2019}, this study exhibits the benefits of learning in providing optimal portfolio allocations.

%%%%%%%%%%%%%%%%%%%%%%%%%%%%%%%%%%%%%%%%%%%%%%%%%%%%%%%%%%%%%%%%%%%%%%%%%%%%%%%%%%%%%%%%%%%%%%%%%%%%%%%%%%%%%%%%%%%%%%%%%%%%%%%%%%%%%%%%%%%%%%%%%%%%%%%%%%%%%%%%%%%%%%%%%%%%%%%%%%%%%%%%%%%%%%%%%%%%%%%%%%%%%%%%%%%%%%%%%%%%%
\section*{Appendix} \label{sec: append}
\addcontentsline{toc}{section}{Appendix}

\subsection{Proof of Proposition \ref{lem: prop_p_bar}} \label{Pr: lem: prop_p_bar}
For all $k = 1, ..., N$, the law under $\overline{\mathbb{P}}$, of $R_k$ given the filtration $\mathcal{G}_{k-1}$ yields the unconditional law under $\mathbb{P}$ of $\e{k}$. Indeed,  since $(\Lambda_k)_k$ is a 
$(\mathbb{P},\Gb)$-martingale, we have from Bayes formula, for all Borelian $F$ $\subset$ $\R^d$, 
\beqs
\overline \P[ R_k \in F |\mathcal{G}_{k-1} ]  &=&  \overline \E[ \mathbbm{1}_{\{R_k \in F \}} |\mathcal{G}_{k-1}]
		\;  =   \; \frac{\E[ \Lambda_k \mathbbm{1}_{\{R_k \in F \}} |\mathcal{G}_{k-1}]}{\E[ \Lambda_k|\mathcal{G}_{k-1}]} \\
		& = & \E[ \Frac{\Lambda_k}{\Lambda_{k-1}}  \mathbbm{1}_{\{R_k \in F \}} |\mathcal{G}_{k-1}] 
		\; =  \;  \E \left[ \frac{g(B+\e{k})}{g(\e{k})} \mathbbm{1}_{\{R_k \in F\}} \big| \mathcal{G}_{k-1} \right] \\
		& =  &\int_{\Rd} \frac{g(B+e)}{g(e)} \mathbbm{1}_{\{B + e  \in F\}} g(e)de 
		\; =  \;  \int_{\Rd} g(z) \mathbbm{1}_{\{z \in F \}} dz \\
		& = & \P[ \e{k} \in F]. 
\enqs
This  means that, under $\overline{\mathbb{P}}$, $R_k$ is independent from $B$ and from $R_1,..,R_{k-1}$ and that $R_k$ has the same probability distribution as $\e{k}$.
\ep

\subsection{Proof of Proposition \ref{lem: unnorm_cond_law}} \label{Pr: lem: unnorm_cond_law}
For any  borelian function $f: \Rd \mapsto \R$ we have, on  one hand, by definition of $\pi_{k+1}$: 
\beqs
\overline{\E} \big[\overline{\Lambda}_{k+1} f(B) | \F{k+1}^o \big] &=& \int_{\Rd}f(b) \pi_{k+1}(db),
\enqs
and, on the other hand, by definition of $\overline{\Lambda}_{k}$: 
\beqs
\overline{\E} [\overline{\Lambda}_{k+1} f(B) | \F{k+1}^o] &=&  \overline{\E} \left[\overline{\Lambda}_{k} f(B) \frac{g(R_{k+1}-B)}{g(R_{k+1})} \middle | \F{k+1}^o \right] \\
	&=& \overline{\E} \left[\overline{\Lambda}_k f(B) g(R_{k+1}-B) \middle | \F{k+1}^o \right]  (g(R_{k+1}))^{-1}\\
	&=& \int_{\Rd}f(b) \frac{g(R_{k+1}-b)}{g(R_{k+1})}\pi_{k}(db),
\enqs
where we use in the last equality the fact that $R_{k+1}$ is independent of $B$ under $\overline{P}$ (recall Proposition \ref{lem: prop_p_bar}). 
By identification, we obtain the expected relation.
\ep
 
\subsection{Proof of Lemma \ref{lem: cond_a}} \label{Pr: lem: cond_a}
Since the support of the probability distribution $\nu$ of $\e{k}$ is  $\Rd$, we notice that  the law of  the random vector 
$Y_k$ $:=$ $e^{R_k} -  \mathbbm{1}_d$ has support equal to $(-1,\infty)^d$. Recall from \eqref{Aqxz} that $a$ $\in$ $A_k^q(x,z)$ iff
\beq \label{eq: ineg_X1_Gauss}
1+  a^\prime Y_{k+1}  
&\geq& 
q  \max\Big[ \frac{z}{x},  1 + a^\prime Y_{k+1} 
\Big], \quad a.s.  
\enq

\vspace{1mm}

\noindent (i) Take some $a$ $\in$ $A_k^q(x,z)$, and  assume that  $a^i$ $<$ $0$ for some $i \in [\![ 1,d]\!]$. Let us then define the event $\Omega_M^i$ $=$ $\{ Y_{k+1}^i \geq M, Y_{k+1}^M \in [0,1], j \neq i\}$, for $M$ $>$ $0$, and observe that 
$\P[\Omega_M^i]$ $>$ $0$.   It follows from \eqref{eq: ineg_X1_Gauss}  that 
\beqs
1 + a_i M + \max_{j\neq i} |a_j| & \geq & q \frac{z}{x}, \quad \mbox{ on } \Omega_M^i, 
\enqs
which leads to a contradiction for $M$ large enough. This shows that $a^i$ $\geq$ $0$ for all  $i \in [\![ 1,d]\!]$, i.e. $A_k^q(x,z)$ $\subset$ $\R_+^d$. 

 \vspace{2mm}
 
 \noindent (ii)  For $\varepsilon$ $\in$ $(0,1)$, let us define the event $\Omega_\varepsilon$ $=$ $\{ Y_{k+1}^i  \leq - 1 + \varepsilon, i =1,\ldots,d\}$, which satisfies  $\P[\Omega_\varepsilon]$ $>$ $0$. 
 For  $a$ $\in$ $A^q(x,z)$, we get from \eqref{eq: ineg_X1_Gauss},  and since $a$ $\in$ $\R_+^d$ by Step (i): 
 \beqs
 1 - (1-\varepsilon) a^\prime \mathbbm{1}_d   & \geq & q \frac{z}{x}, \quad \mbox{ on } \Omega_\varepsilon.  
 \enqs
 By taking $\varepsilon$ small enough, this shows by a contradiction argument that 
 \beq  \label{Aq=} 
 A_k^q(x,z)  & \subset & \Big\{ a \in \R_+^d:  1 - a^\prime \mathbbm{1}_d   \geq q \frac{z}{x} \Big\}. \; = : \; \tilde A^q(x,z).
 \enq
  
 \noindent (iii) Let us finally check the equality in \eqref{Aq=}. Fix some $a$ $\in$ $\tilde A^q(x,z)$.  Since the random vector $Y_{k+1}$ is valued in $(-1,\infty)^d$,  it is clear that 
 \beqs
 1+  a^\prime Y_{k+1}   & \geq & 1 - a^\prime \mathbbm{1}_d  \; \geq \; q \frac{z}{x}  \; \geq \; 0,  \quad a.s.,
 \enqs
 and thus 
 \beqs
 1+  a^\prime Y_{k+1}   & \geq & q \big[ 1+  a^\prime Y_{k+1} \big],  \quad a.s., 
 \enqs
 which proves \eqref{eq: ineg_X1_Gauss}, hence the equality $A^q(x,z)$ $=$  $\tilde A(x,z)$. 
\ep

\subsection{Proof of Lemma \ref{lem: Aq_gen}} \label{Pr: lem: Aq_gen}
\noindent  {\bf 1.}  Fix  $q_1 \leq q_2$ and $(x,z)$ $\in$ $\Sc^{q_2}$ $\subset$ $\Sc^{q_1}$. We then have 
\beqs
a \in A^{q_2}{}(x,z)  \; \Rightarrow \; a \in \Rd_+ \mbox{ and }  a^\prime \mathbbm{1}_d \; \leq \;  1 - q_2\frac{z}{x} \; \leq \;  1 - q_1\frac{z}{x} \; \implies \;  a \in A^{q_1}{}(x,z), 
\enqs
which  means that $A^{q_2}{}(x,z) \subseteq A^{q_1}{}(x,z)$. 

\noindent  {\bf 2.} Fix $q$ $\in$ $(0,1)$, and consider the decreasing sequence $q_n = q + \frac{1}{n}$, $n$ $\in$ $\N^*$.  For any $(x,z)$ $\in$ $\Sc^{q_n}$, we then have 
$A^{q_n}(x,z)$ $\subseteq$ $A^{q_{n+1}}(x,z)$  $\subset$ $A^a(x,z)$, which implies that the sequence of increasing sets $A^{q_n}(x,z)$ admits a limit equal to
\beqs
\lim_{n \to \infty} A^{q_n}(x,z)  &=&  \underset{n \geq 1}{\cup} A^{q_n}(x,z) \; = \;  A^q(x,z),
\enqs
since $\lim_{n \to \infty} q_n = q$. This shows the right continuity of $q$ $\mapsto$ $A^q(x,z)$.  Similarly, by considering  the increasing sequence $q_n = q - \frac{1}{n}$, $n$ $\in$ $\N^*$, 
we see that  for any $(x,z)$ $\in$ $A^q(x,z)$, the sequence of decreasing sets $A^{q_n}(x,z)$ admits a limit equal to 
\beqs
\lim_{n \to \infty} A^{q_n}(x,z)  &=&  \underset{n \geq 1}{\cap} A^{q_n}(x,z) \; = \;  A^q(x,z),
\enqs
since $ \lim_{n \to \infty} q_n = q$. This proves the continuity in $q$ of the set $A^q(x,z)$. 

\noindent  {\bf 3.} Fix $q$ $\in$ $(0,1)$, and $(x_1,z)$, $(x_2,z)$ $\in$ $\Sc^q$ s.t. $x_1\leq x_2$. Then, 
\beqs
a \in   A^{q}{}(x_1,z) \;  \implies \;  a \in \Rd_+ \mbox{ and }  a^\prime \mathbbm{1}_d \leq 1-q\frac{z}{x_1} \leq 1-q\frac{z}{x_2} \; \implies \;  a \in A^{q}{}(x_2,z),
\enqs
which shows that  $A^{q}{}(x_1, z) \subseteq A^{q}{}(x_2, z)$. 

\vspace{3mm}

\noindent  {\bf 4.}  Fix $q$ $\in$ $(0,1)$, $(x,z)$ $\in$ $A^a(x,z)$. Then, for any  $a_1, a_2$ of the set $A^{q}{}(x,z)$, and $\beta$ $\in$ $(0,1)]$, and 
denoting by $a_3$ $=$ $\beta a_1 + (1- \beta) a_2$ $\in$ $\R_+^d$, we have  
\beqs
a_3^\prime \mathbbm{1}_d  \; = \;   \beta a_1^\prime \mathbbm{1}_d + (1- \beta) a_2^\prime \mathbbm{1}_d &\leq &  
\beta  \big(1-q\frac{z}{x} \big) + (1- \beta) \big(1-q\frac{z}{x}\big) \; = \;  1-q\frac{z}{x}.
\enqs
This proves the convexity of the set  $A^{q}{}(x, z)$.

\noindent   {\bf 4.} The homogeneity property of $A^q(x,z)$ is obvious from its very definition.  
\ep

\subsection{Proof of Lemma \ref{lem: homo_value_function}} \label{Pr: lem: homo_value_function}
We prove the result by backward induction on time $k$  from the dynamic programming equation for the value function.  

\vspace{1mm}

\noindent $\bullet$ At time $N$, we have  for all $\lambda$ $>$ $0$, 
\beqs
v_N(\lambda x, \lambda z,\mu) &=& \frac{(\lambda x)^p}{p}  = \lambda^p v_N( x,z,\mu), 
\enqs
which shows the required homogeneity property. 

\noindent $\bullet$ Now, assume that the homogeneity property holds at time $k+1$, i.e $v_{k+1}(\lambda x,\lambda z, \mu)$ $=$ $\lambda^{p} v_{k+1}(x,z,\mu)$ for any $\lambda$ $>$ $0$. 
Then, from the backward relation \eqref{eq: dyn_prog_gen}, and the homogeneity property of $A^q(x,z)$ in Lemma \ref{lem: Aq_gen},  it is clear that $v_k$ inherits from $v_{k+1}$ the 
homogeneity property.  
\ep

\subsection{Proof of Lemma \ref{lem: v_prop}} \label{Pr: lem: v_prop}
\noindent {\bf 1.} We first show by backward induction that $r \mapsto w_{k}(r, \cdot$) is nondecreasing in on $[q,1]$ for all $k \in [\![0,N ]\!]$.\\ 
\noindent $\bullet$ For any $r_1,r_2$ $\in$ $[q,1]$, with $r_1\leq  r_2$,  and $\mu$ $\in$ $\mathcal{M}_+$, we have at time $N$
\beqs
w_N(r_1,\mu) \; = \;  U(r_1) \mu(\R^d) &\leq&  U(r_2) \mu(\R^d)  \; = \;  w_N(r_2,\mu).
\enqs 
This shows that $w_N(r,\cdot)$ is nondecreasing on $[q,1]$.  

\noindent $\bullet$ Now, suppose by  induction hypothesis that  $r$ $\mapsto$ $w_{k+1}(r,\cdot)$ is nondecreasing. Denoting by $Y_k$ $:=$ $e^{R_k} -  \mathbbm{1}_d$ the random vector valued in 
$(-1,\infty)^d$, we see that for all $a$ $\in$ $A^q(r_1)$ 
\beqs
%x_1 \big( 1 + a^\prime Y_{k+1}  \big) &\leq&  x_2 \big( 1 + a^\prime Y_{k+1} \big), \quad a.s.  \\
\min\Big[1,r_1 \big( 1 + a^\prime Y_{k+1}  \big) \Big] & \leq & \min\Big[1,r_2 \big( 1 + a^\prime Y_{k+1}  \big) \Big],  \quad a.s. 
\enqs
since $1 + a^\prime Y_{k+1}$ $\geq$ $1-a^\prime\mathbbm{1}_d$ $\geq$ $q\frac{1}{r_1}$ $\geq$ $0$. Therefore, from backward dynamic programming Equation \eqref{eq: dyn_prog_CRRA}, 
and noting that $A^q(r_1)$ $\subset$ $A^q(r_2)$, we have  
\beqs
w_k(r_1,\mu) &=&  \Sup_{a \in A^q_(r_1)}  \overline\E\Big[ w_{k+1} \big(    
\min\big[1,   r_1\big(1 +  a^\prime Y_{k+1} \big) \big], \bar g(R_{k+1}-\cdot) \mu \big)  \Big] \\
& \leq & \Sup_{a \in A^q(r_2)}  \overline\E\Big[ w_{k+1} \big(    
\min\big[1,   r_2\big(1 +  a^\prime Y_{k+1} \big) \big], \bar g(R_{k+1}-\cdot) \mu \big)  \Big]  \; = \; w_k(r_2,\mu), 
\enqs
which shows the required nondecreasing property at time $k$.\\

\noindent {\bf 2.}
We prove the concavity  of $r$ $\in$ $[q,1]$  $\mapsto$ $w_k(r,\cdot)$  by backward induction for all $k \in [\![0,N ]\!]$. For $r_1,r_2$ $\in$ $[q,1]$, and $\lambda$ $\in$ $(0,1)$, we set 
 $r = \lambda r_1 + (1-\lambda)r_2$, and for $a_1 \in A^q(r_1)$, $a_2 \in A^q(r_2)$, we set $a$ $=$ $\big( \lambda r_1 a_1+ (1-\lambda)r_2 a_2 \big)/r$ 
 which belongs to $A^q(r)$. Indeed, since $a_1, a_2 \in \Rd_+$, we have  $a \in \Rd_+$, and
\beqs
a  \; = \;  \Big( \frac{\lambda r_1 a_1+ (1-\lambda) r_2 a_2 }{r}\Big)^\prime \mathbbm{1}_d  &\leq& 
\frac{\lambda r_1}{r}\big(1- \frac{q}{r_1}\big) +\frac{(1-\lambda)r_2}{r}\big(1- \frac{q}{r_2}\big) \; = \;  1 - \frac{q}{r}.
\enqs 
\noindent $\bullet$ At time $N$, for fixed $\mu \in \mathcal{M}_+$, we have
\beqsal
w_N \big(\lambda r_1 + (1-\lambda) r_2,\mu\big) &= \;  U(\lambda r_1 + (1-\lambda) r_2) \\
&\geq \;  \lambda U(r_1) + (1-\lambda)U(r_2) \;= \;  \lambda w_N(r_1,\mu) + (1-\lambda) w_N(r_2, \mu),
\enqsal
since $U$ is concave. This shows that $w_N(r,\cdot)$ is concave on $[q,1]$.

\noindent $\bullet$ Suppose now the induction hypothesis holds true at time $k+1$: $w_{k+1} (r, \cdot)$ is concave on $[q,1]$. From 
the backward dynamic programming relation \eqref{eq: dyn_prog_CRRA}, we then have 
\begin{eqnarray*}
& & \lambda w_k(r_1,\mu) + (1- \lambda) w_k(r_2,\mu) \\
& \leq & \lambda \E\Big[ w_{k+1} \big(  \min[1, r_1(1 + a_1^\prime Y_{k+1})], \bar g(R_{k+1}-\cdot)\mu \big) \Big]  \\ 
& &  \hspace{3cm} + (1-\lambda)\E \Big[ w_{k+1} \big(\min[1, r_2(1 + a_2^\prime Y_{k+1})], \bar g(R_{k+1}-\cdot)\mu \big) \Big] \\
& \leq &  \E\Big[ w_{k+1} \big( \lambda \min[1, r_1(1 + a_1^\prime Y_{k+1})]  + (1-\lambda) \min[1, r_2(1 + a_2^\prime Y_{k+1})], \bar g(R_{k+1}-\cdot) \mu\big)\Big] \\
& = &  \E\Big[ w_{k+1} \big( \min[1,r(1 + a^\prime Y_{k+1})],\bar g(R_{k+1}-\cdot) \mu \big)\Big]  \;   \leq \;  w_k(r,\mu),
\end{eqnarray*}
where we used for the second inequality, the induction hypothesis joint with the concavity of  $x \mapsto \min(1,x)$, and the nondecreasing monotonicity of   
$r$ $\mapsto$ $w_{k+1}(r,\cdot )$.  This shows the required inductive concavity property of $r \mapsto w_k(r, \cdot)$ on $[q,1]$.
\ep

%%%%%%%%%%%%%%%%%%%%%%%%%%%%%%%%%%%%%%%%%%%%%%%%%%%%%%%%%%%%%%%%%%%%%%%%%%%%%%%%%%%
%%%%%%%%%%%%%%%%%%%%%%%%%%%%%%% BIBLIOGRAPHY %%%%%%%%%%%%%%%%%%%%%%%%%%%%%%%%%%%%%%
%%%%%%%%%%%%%%%%%%%%%%%%%%%%%%%%%%%%%%%%%%%%%%%%%%%%%%%%%%%%%%%%%%%%%%%%%%%%%%%%%%%
\bibliographystyle{chicagoa}

\bibliography{Bib_Projet}{}

%%%%%%%%%%%%%%%%%%%%%%%%%%%%%%%%%%%%%%%%%%%%%%%%%%%%%%%%%%%%%%%%%%%%%%%%%%%%%%%%%%%

\end{document}